\documentclass{pasj01}
\usepackage[switch,mathlines]{lineno}
\Received{$\langle$2023-11-23$\rangle$}
\Accepted{$\langle$2024-02-09$\rangle$}
\Published{$\langle$publication date$\rangle$}
\SetRunningHead{Sub-mm Water Maser in NGC 1052}{NGC 1052}
\begin{document}

\title{Sub-Parsec-Scale Jet-Driven Water Maser with Possible Gravitational Acceleration in the Radio Galaxy NGC 1052}
\author{Seiji \textsc{Kameno},\altaffilmark{1,2,3} %
        Yuichi \textsc{Harikane},\altaffilmark{4} %
        Satoko \textsc{Sawada-Satoh},\altaffilmark{5}
        Tsuyoshi \textsc{Sawada},\altaffilmark{1,2} %
        Toshiki \textsc{Saito},\altaffilmark{2} %
        Kouichiro \textsc{Nakanishi},\altaffilmark{2,3} %
        Elizabeth \textsc{Humphreys},\altaffilmark{1,6} and %
        C. M. Violette \textsc{Impellizzeri}\altaffilmark{7} %
}

\altaffiltext{1}{Joint ALMA Observatory, Alonso de C\'{o}rdova 3107 Vitacura, Santiago 763-0355, Chile}
\altaffiltext{2}{National Astronomical Observatory of Japan, 2-21-1 Osawa, Mitaka, Tokyo 181-8588, Japan}
\altaffiltext{3}{Department of Astronomy, School of Science, Graduate University for Advanced Studies (SOKENDAI), Tokyo 181-8588, Japan}
\altaffiltext{4}{Institute for Cosmic Ray Research, The University of Tokyo, 5-1-5 Kashiwanoha, Kashiwa, Chiba 277-8582, Japan}
\altaffiltext{5}{Graduate School of Science, Osaka Metropolitan University, 1-1 Gakuen-cho, Naka-ku, Sakai, Osaka, 599-8531, Japan}
\altaffiltext{6}{European Southern Observatory, Karl-Schwarzschild-Strasse 2 D-85748, Garching, Germany}
\altaffiltext{7}{Leiden Observatory, Leiden University, PO Box 9513, 2300 RA, Leiden, The Netherlands}

\email{seiji.kameno@alma.cl}


\maketitle

\begin{abstract}
We report sub-pc-scale observations of the 321-GHz H$_2$O emission line in the radio galaxy NGC 1052.
The H$_2$O line emitter size is constrained in $< 0.6$ milliarcsec distributed on the continuum core component.
The brightness temperature exceeding $10^6$ K and the intensity variation indicate certain evidence for maser emission.
The maser spectrum consists of redshifted and blueshifted velocity components spanning $\sim 400$ km s$^{-1}$, separated by a local minimum around the systemic velocity of the galaxy.
Spatial distribution of maser components show velocity gradient along the jet direction, implying that the population-inverted gas is driven by the jets interacting with the molecular torus.
We identified significant change of the maser spectra between two sessions separated by 14 days.
The maser profile showed a radial velocity drift of $127 \pm 13$ km s$^{-1}$ yr$^{-1}$ implying inward gravitational acceleration at 5000 Schwarzschild radii.
The results demonstrate feasibility of future VLBI observations to resolve the jet-torus interacting region.

\end{abstract}

%\pagewiselinenumbers
\section{Introduction}\label{sec:introduction}
The Atacama Large Millimeter/submillimeter Array (ALMA) detected H$_2$O $J_{Ka, Kc} = 10_{2,9} - 9_{3,6}$ emission at rest frequency of 321.225677 GHz in the radio galaxy, NGC 1052 \citep{2023PASJ...75L...1K}.
This is the first submillimeter maser detection in a radio galaxy and the most luminous 321-GHz H$_2$O maser known to date with the isotropic luminosity of 1090 L$_{\Sol}$.
The most plausible interpretation is maser amplification of background synchrotron emission through population-inverted molecular gas in a torus.
However, the angular resolution of $0^{\prime \prime}.68 \times 0^{\prime \prime}.56$ with a compact array configuration up to 500 m was insufficient to rule out the possibility of thermal emission.
A higher angular resolution was required to confirm for maser emission by brightness temperature ($T_{\rm B}$), which is a criterion to classify the emission mechanism by comparing with the upper energy level of $E_u / k_b = 1861$ K, where $k_b$ is the Boltzmann constant.
\citet{2023PASJ...75L...1K} mentioned that baseline length of $> 3500$ m would provide concrete evidence for non-thermal emission.
New ALMA long-baseline observations were deemed necessary to confirm that the emission was of masers, and also allows us to clarify the spatial distribution of population-inverted molecular gas.
%ALMA long-baseline configuration delivers the most straightforward confirmation for maser emission, and also allows us to clarify the spatial distribution of population-inverted molecular gas.

NGC 1052 is a unique target emanating well-collimated \citep{2020AJ....159...14N, 2022A&A...658A.119B} double-sided sub-relativistic jets with a bulk speed of $0.26c - 0.53c$ \citep{2003A&A...401..113V, 2019A&A...623A..27B}, where $c$ is the speed of light.
Kinematic studies clarified that eastern and western jets reside approaching and receding sides, respectively, with the viewing angle $> 57^{\circ}$ \citep{2003A&A...401..113V} under assumption of intrinsic symmetry.
\citet{2022A&A...658A.119B} questioned the symmetry and confirmed the same sign of the viewing angle.
A multi-phase torus in pc-scale vicinity of the core is seen nearly edge-on and has been investigated by various probes such as molecular absorption lines \citep{2002A&A...381L..29O, 2004A&A...428..445L, 2008evn..confE..33I, 2016ApJ...830L...3S, 2019ApJ...872L..21S, 2020ApJ...895...73K, 2023ApJ...944..156K}, H\emissiontype{I} absorption line \citep{1983A&A...119L...3S, 2003A&A...401..113V}, H$_2$O 22-GHz maser emission \citep{1994ApJ...437L..99B, 1998ApJ...500L.129C, 2003ApJS..146..249B, 2005ApJ...620..145K, 2008ApJ...680..191S}, and free--free absorption (FFA) in thermal plasma \citep{2001PASJ...53..169K, 2003PASA...20..134K, 2003A&A...401..113V, 2004A&A...426..481K}.

H$_2$O megamaser at 22 GHz is a firm probe for sub-pc-scale molecular gas in AGNs and its (sub-)Keplerian rotation curve allows us to weigh a central supermassive black hole (SMBH), to study the molecular gas distribution at the highest resolutions around and AGN, and to measure geometrical distance independently from other estimation \citep{1993Natur.361...45N, 1995Natur.373..127M, 1995PNAS...9211427M, 1995Natur.378..697K, 2005ApJ...629..719H, 2008ApJ...672..800H, 2013ApJ...775...13H, 2022NatAs...6..885I, 2023ApJ...951..109G}.
The reach of 22-GHz H$_2$O maser as a cosmological distance ladder is constrained by an angular resolution to clarify the rotation curve.

Since the first detection of the extragalactic 183-GHz and (tentative) 439-GHz H$_2$O masers in NGC 3079 \citep{2005ApJ...634L.133H}, submillimeter masers have become recognized as new and important probes for AGN.
\citet{2013ApJ...768L..38H} searched for the 321-GHz H$_2$O maser in five type-2 Seyfert galaxies and detected in the Circinus galaxy.
That is followed by the second detection in NGC 4945 \citep{2016ApJ...827...68P, 2016ApJ...827...69H, 2021ApJ...923..251H} and the first detection in a radio galaxy NGC 1052 \citep{2023PASJ...75L...1K}.
With a higher angular resolution by $14 \times$ with the same baseline length, the 321-GHz H$_2$O maser offers potential to extend the reach of the cosmological distance ladder.

The 321-GHz maser requires a condition with higher temperature and density compared with that of the 22-GHz maser.
\citet{1991ApJ...368..215N} modeled the excitation condition for the 321-GHz H$_2$O maser under a physical temperature of $T_k = 1000$ K and found that the maser emissivity was maximized with a density of hydrogen nuclei $\sim 10^9$ cm$^{-3}$.
\citet{1997MNRAS.285..303Y} also concluded that the population inversion condition for the 321-GHz H$_2$O maser was found at $T_k > 1000$ K for $n($H$_2) = 8 \times 10^8$ and $4\times 10^9$ cm$^{-3}$ with $n($H$_2$O$) = 10^4$ cm$^{-3}$.
\citet{2016MNRAS.456..374G} modeled the excitation conditions for possible H$_2$O maser transitions for evolved star environments and clarified that the population-inversion for the 321-GHz maser requires $T_k > 1000$ K and $n($H$_2$O$) > 10^4$ cm$^{-3}$.
Thus, we expect closer distance to the SMBH, greater gravitational acceleration, wider velocity range, and faster proper motion of disk rotation.

In this paper, we report the results from ALMA observations in the long-baseline array configuration (C-10) for 321-GHz H$_2$O emission toward NGC 1052.
This paper is organized as follows: Section \ref{sec:method} describes observation conditions of ALMA and data reduction procedures. Section \ref{sec:results} presents results of continuum images, spectral profiles, size and position of velocity components of the 321-GHz H$_2$O emission.
In section \ref{sec:discussion} we develop discussion about justification of maser emission, velocity drift, velocity gradient, jet-torus interaction, and torus structure. Then we summarize our conclusions in section \ref{sec:summary}.
We employ the systemic velocity of $V_{\rm sys, LSR} = 1492$ km s$^{-1}$, the luminosity distance of $D_L = 17.6$ Mpc, the angular distance of $D_A = 17.5$ Mpc, and the linear scale of $85$ pc arcsec$^{-1}$ \citep{2020ApJ...895...73K}.

\section{Methods}\label{sec:method}

\subsection{Observations}
\begin{figure}
 \begin{center}
 \includegraphics[width=0.475\linewidth]{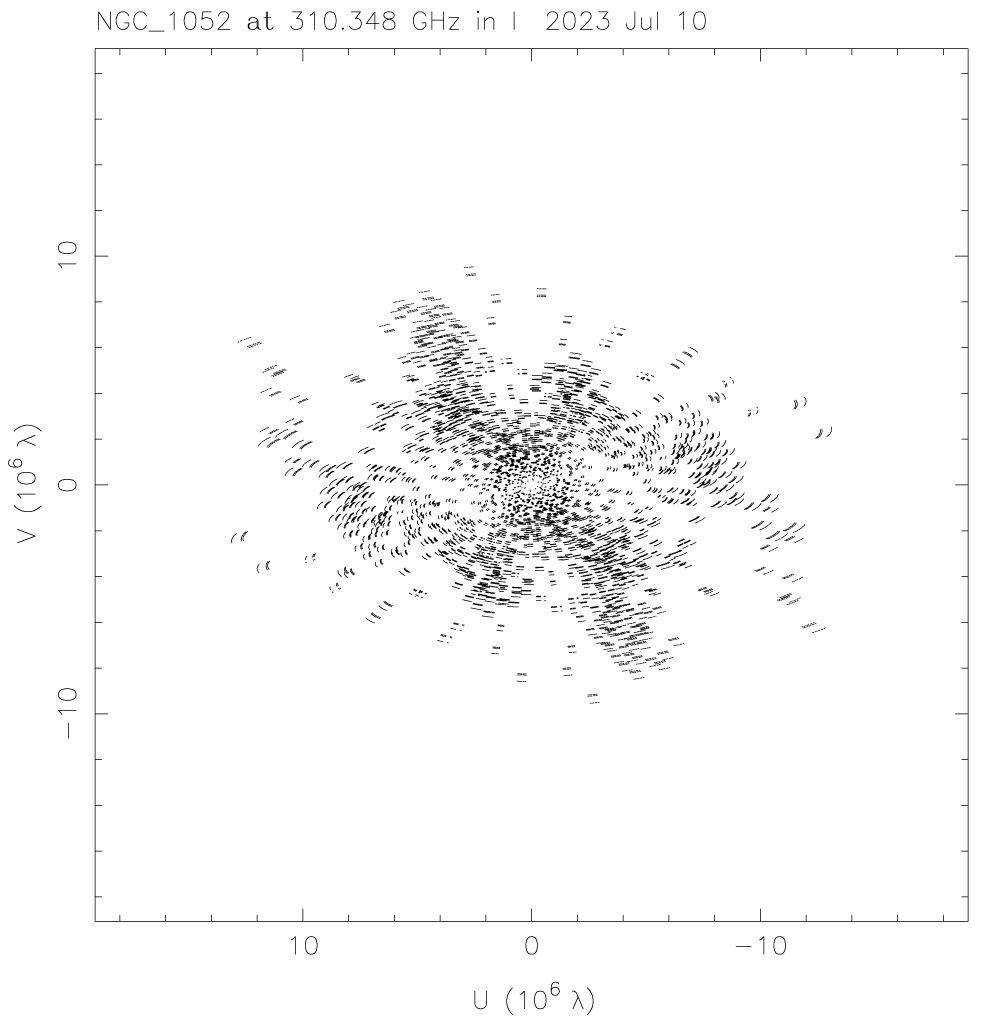}
 \includegraphics[width=0.475\linewidth]{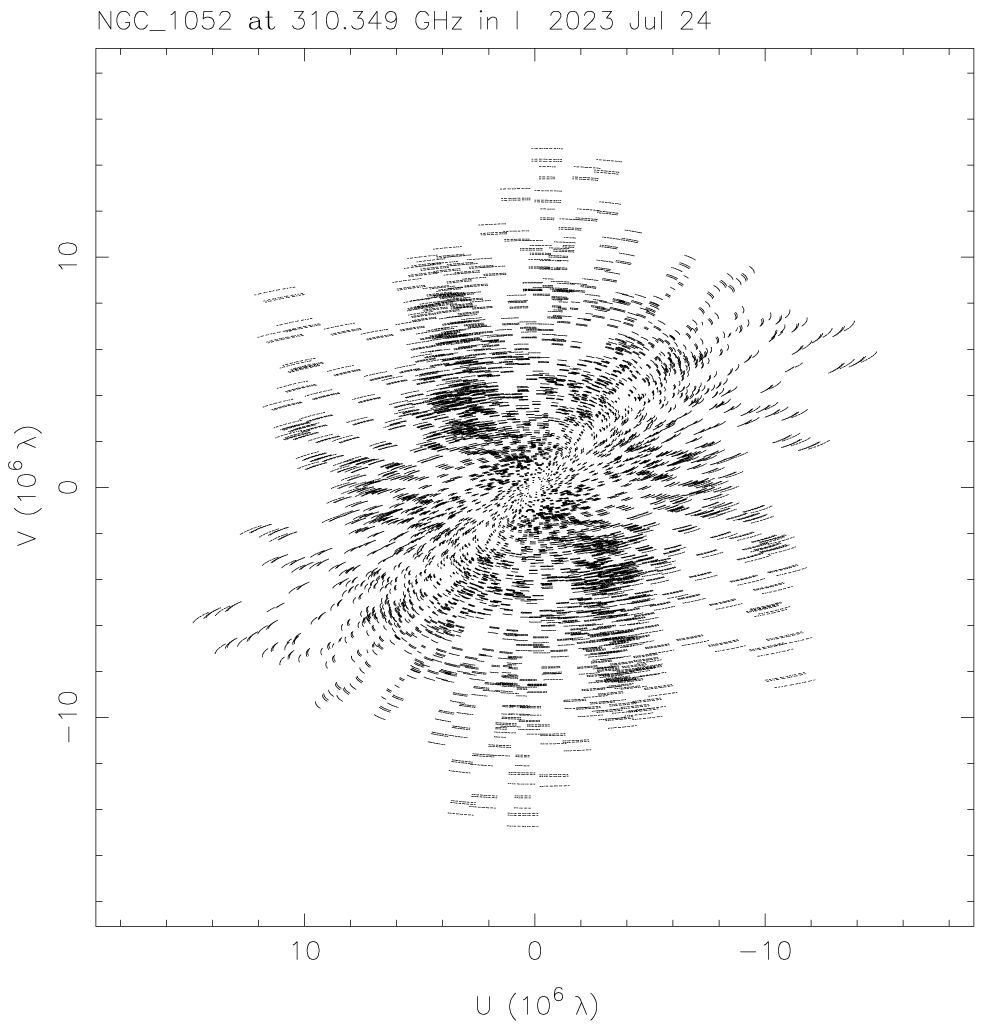}
 \end{center}
 \caption{Distributions of spatial frequencies, also known as $(u, v)$ coverages, obtained on 2023-07-10 (left) and 2023-07-24 (right).
}\label{fig:uvcover}
\end{figure}

ALMA observations under C-10 (the most extended) array configuration with the maximum baseline length of 15.2 km have been carried out twice on 2023-07-10 and 2023-07-24 as summarized in table \ref{tab:obslog}.
Integration times on the target (NGC 1052) and on a bandpass calibrator were 1108 s and 912 s, respectively, for each execution.
The first execution, uid://A002/X109d26e/X12c34, was significantly affected by high wind speed that made three antennas shutdown in the beginning and then four antennas later. Absence of antennas in the southern arm degraded the angular resolution in north-south direction poorer than the requested resolution of $0^{\prime \prime}.015$.
The retake, uid://A002/X10a7a20/X65a8, was successful and satisfied the required resolution.
Median values of the system noise temperatures were 139 K and 132 K under the condition with the precipitable water vapor (PWV) of 0.58 mm and 0.48 mm, respectively.
The distribution of acquired spatial frequencies, also known as $(u, v)$ coverage, is shown in figure \ref{fig:uvcover}.

We configured four spectral windows (SPWs) for four 2-GHz basebands (BBs), assigning two in the upper sideband (USB) and two in the lower sideband (LSB).
BB4 in USB was tuned to cover the H$_2$O $J_{Ka, Kc} = 10_{2,9} - 9_{3,6}$ emission and the SPW was set in the frequency division mode (FDM) with a channel separation of 976.562 kHz corresponding to 0.914 km s$^{-1}$.
Other three SPWs were set in the time division mode (TDM) to maximize the sensitivity in continuum emission with a effective bandwidth of 5.4 GHz.

%-------- Observation Log Table
\begin{table*}
  \tbl{Observation logs and performances.}{%
  \begin{tabular}{lllllrr}
      \hline
      ExecBlock UID & Date & $N_{\rm ant}$  & Bandpass & Synthesized beam         & \multicolumn{2}{c}{Image rms (mJy beam$^{-1}$)} \\ 
	  uid://A002/   & (year-month-day)      &  &  & major $\times$ minor, PA & continuum    & line  \\
	  (1) & (2) & (3) & (4) & (5) & (6) & (7) \\
      \hline
      X109d26e/X12c34 & 2023-07-10 & 37\footnotemark[$*$] & J0423-0120 & $0^{\prime \prime}.017 \times 0^{\prime \prime}.011 $, $-7^{\circ}.4$  & 0.57 & 2.3 \\
      X10a7a20/X65a8  & 2023-07-24 & 50                   & J0238+1636 & $0^{\prime \prime}.012 \times 0^{\prime \prime}.010 $, $41^{\circ}.5$  & 0.24 & 2.0 \\
      \hline
    \end{tabular}}\label{tab:obslog}
\begin{tabnote}
(1) Execution block UID. (2) Date of observation. (3) Number of antennas. (4) Bandpass calibrator selected by the operation software. (5) Major axis, minor axis, and position angle of the synthesized beam. (6) Image rms of continuum with the bandwidth of 5.4 GHz. (7) Image rms with a spectral resolution of 976.562 kHz.

\footnotemark[$*$]Among 44 antennas, three antennas were lost before the first scan of NGC 1052 due to high wind speed. Additionally four antennas were lost in the last 21 minutes. \\ 
%\footnotemark[$\dag$] Explanation of value 3. 
%\footnotemark[$\ddag$] \\ 
%\footnotemark[$\S$]  ... \\ 
%\footnotemark[$\|$]  ... \\
%\footnotemark[$\sharp$]  ... \\  
%\footnotemark[$**$]  ... \\ 
%\footnotemark[$\dag\dag$]  ... \\ 
\end{tabnote}
\end{table*}

\subsection{Data Reduction}
Reduction scripts and reduced data are available in the GitHub repository.\footnote{https://github.com/kamenoseiji/ALMA-2022.A.00023.S}

Phase and amplitude calibrations were applied following the standard way, using CASA \citep{2022PASP..134k4501C} 6.4.1.
We used bandpass calibrators, listed in table \ref{tab:obslog}, as a flux calibrator assuming 2.104 Jy for J$0423-0120$ and 1.304 Jy for J$0238+1636$.
This amplitude calibration yields $\sim 3$\% uncertainty.
We applied smoothed bandpass calibration \citep{2012PASJ...64..118Y} with a 9-channel smoothing width to improve the signal-to-noise ratio (SNR) in the spectra.
The target, NGC 1052, was bright and compact enough to allow phase calibration by itself.

\begin{figure*}
 \begin{center}
 \includegraphics[width=\linewidth]{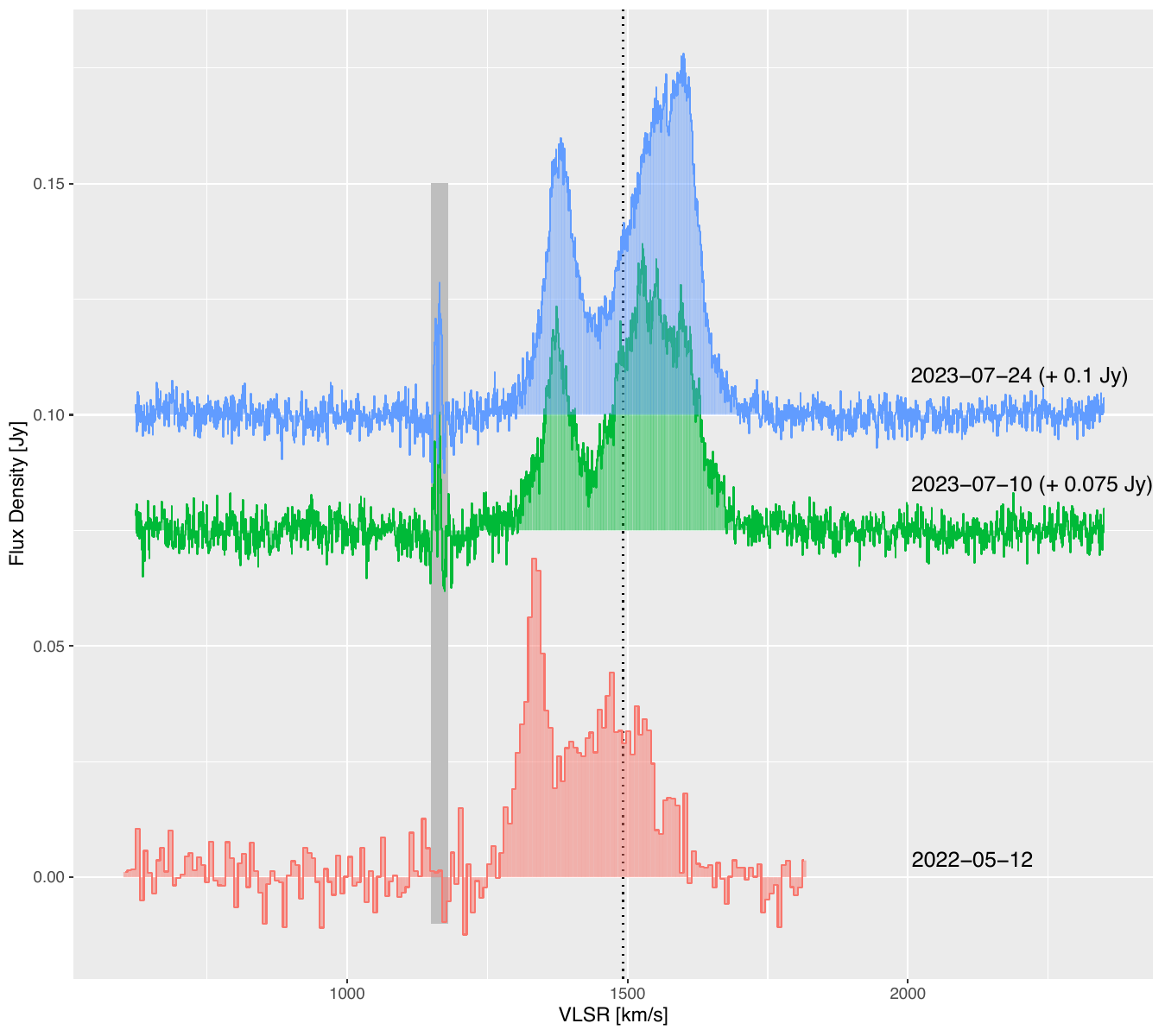}
 \end{center}
 \caption{Continuum-subtracted spectra of H$_2$O $J_{Ka, Kc} = 10_{2,9} - 9_{3,6}$ emission in NGC 1052.
The bottom red spectrum was taken from \citet{2023PASJ...75L...1K} observed on 2022-05-12 with a coarser spectral resolution of 7.8125 MHz.
The middle green and top blue spectra stand for uid://A002/X109d26e/X12c34 on 2023-07-10 and uid://A002/X10a7a20/X65a8 on 2023-07-24, offset by 0.075 Jy and 0.1 Jy, respectively.
The dotted vertical line indicates the systemic velocity of 1492 km s$^{-1}$.
The gray strip around 1150 km s$^{-1}$ masks the telluric ozone feature at 319.9 GHz.
}\label{fig:maser_spectrum}
\end{figure*}

We produced continuum-subtracted cross power spectra in BB4.
Then, we used the task {\tt tclean} with the natural weighting to produce image cubes.
Spectral profiles of the H$_2$O emission line, shown in figure \ref{fig:maser_spectrum}, were sampled at the single phase-center pixel.
The continuum-subtracted visibilities were also used to estimate sizes of the emission region and to measure positions of emission, to be presented in section \ref{sec:results}.

We performed continuum imaging with Difmap \citep{1994BAAS...26..987S}, using channel-averaged visibilities of BB 1--3.
We employed uniform weighting that yielded the synthesized beams of $0^{\prime \prime}.017 \times 0^{\prime \prime}.011$ and $0^{\prime \prime}.012 \times 0^{\prime \prime}.010$, respectively, as stated in table \ref{tab:obslog}.

After subtracting a point CLEAN component at the center position, significant emission remained as an elongated structure spanning $\sim 0^{\prime \prime}.001$ in PA$\sim 70^{\circ}$. The structure was decomposed by CLEAN components aligned in the same PA.
While self calibration with a single central component yields the agreement factors\footnote{The agreement factor in Difmap self calibration is equivalent to $\sqrt{\chi^2_N}$, where $\chi^2_N$ does not take account of the number of degree of freedom.} of $\sigma = 3.94$ and $4.93$ for 2023-07-10 and 2023-07-24, respectively, those with the aligned CLEAN components result in $\sigma = 3.57$ and $1.81$.
Significant improvements in agreement factors justify the reality of the elongated structure.
The elongated structure is buried by the central component if the CLEANed images are restored with the main lobe of the synthesized beams.
To highlight the structure, we produced super-resolution images with a 1.5-milliarcsec (mas) circular restoring beam.

\section{Results}\label{sec:results}
Table \ref{tab:obslog} summarizes imaging performance of the observations.
Image rms is estimated by statistics in emission-free area of syhthesized images.
While rms in spectral-line channel map was dominated by thermal noise, that in continuum image involves systematic errors such as sidelobe leakages.

\subsection{Line profiles}
Continuum-subtracted spectra are shown in figure \ref{fig:maser_spectrum}.
The weighted mean velocities were 1502.8 km s$^{-1}$ and 1507.5 km s$^{-1}$ on 2023-07-10 and 2023-07-24, respectively.
The integrated flux densities of $10.79 \pm 0.33$ Jy km s$^{-1}$ and $14.32 \pm 0.43$ Jy km s$^{-1}$ correspond to the isotropic luminosities of $1120 \pm 34$ L$_{\Sol}$ and $1486 \pm 45$ L$_{\Sol}$, respectively.

The profile mainly consisted of two velocity components --- the blueshifted narrow component and the redshifted wide component --- separated by a local minimum near the systemic velocity.
The velocity range of 1280 -- 1580 km s$^{-1}$ on 2022-05-12 was shifted redward to 1300 -- 1700 km s$^{-1}$
The local minima between red and blue components also redshifted in later observations.
The velocity shifts recall the velocity drift of the systemic velocity component of the 22-GHz H$_2$O maser in NGC 4258 \citep{1994ApJ...437L..35H,1995A&A...304...21G,1995PASJ...47..771N}.
We will discuss about the velocity drift in section \ref{subsec:acceleration}.

While the blue component showed a simple single-peaked profile, the red component contained multiple narrow sub-components.
The peak intensity of the red component was lower than that of the blue one on 2022-05-12, but it got brighter exceeding the blue one in later epochs.

\begin{figure*}
 \begin{center}
  \includegraphics[width=0.45\linewidth]{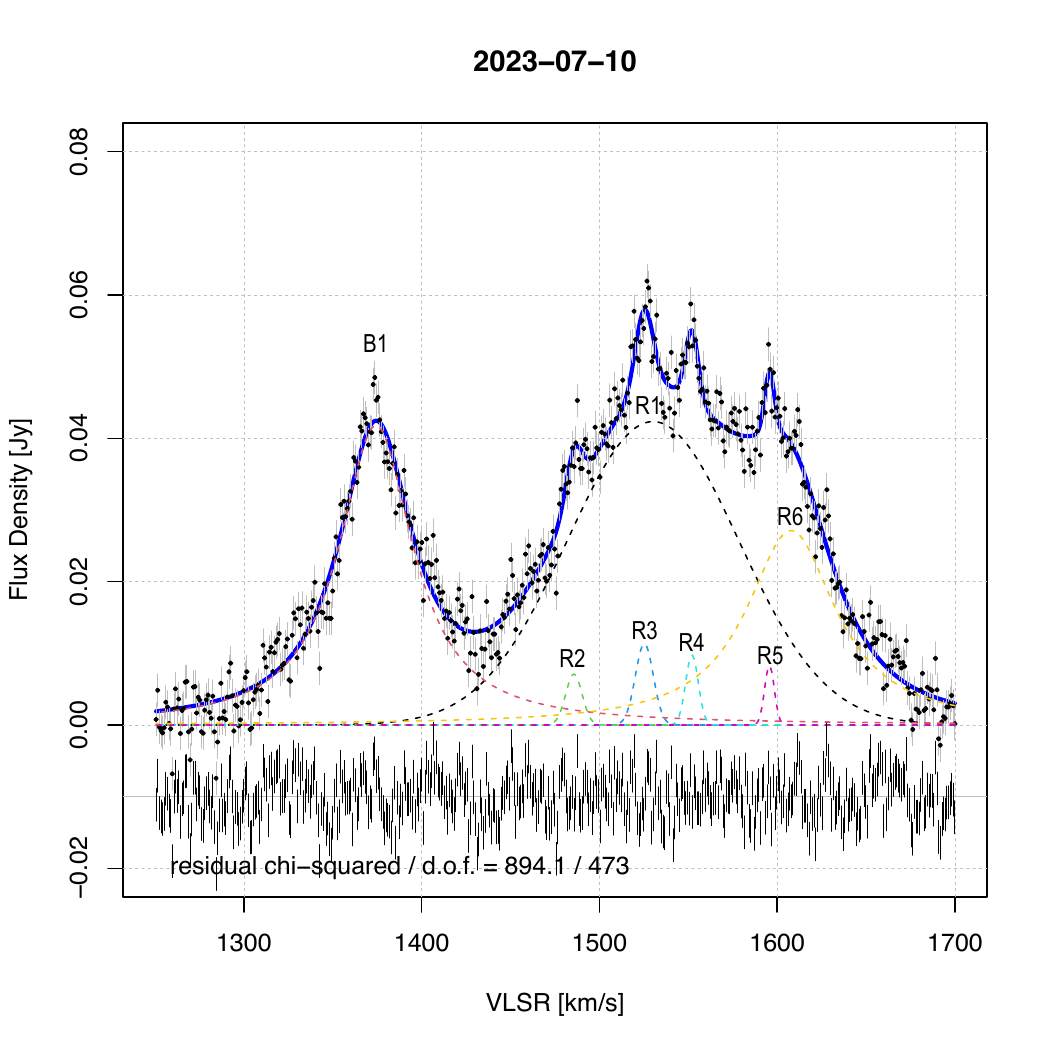}
  \includegraphics[width=0.45\linewidth]{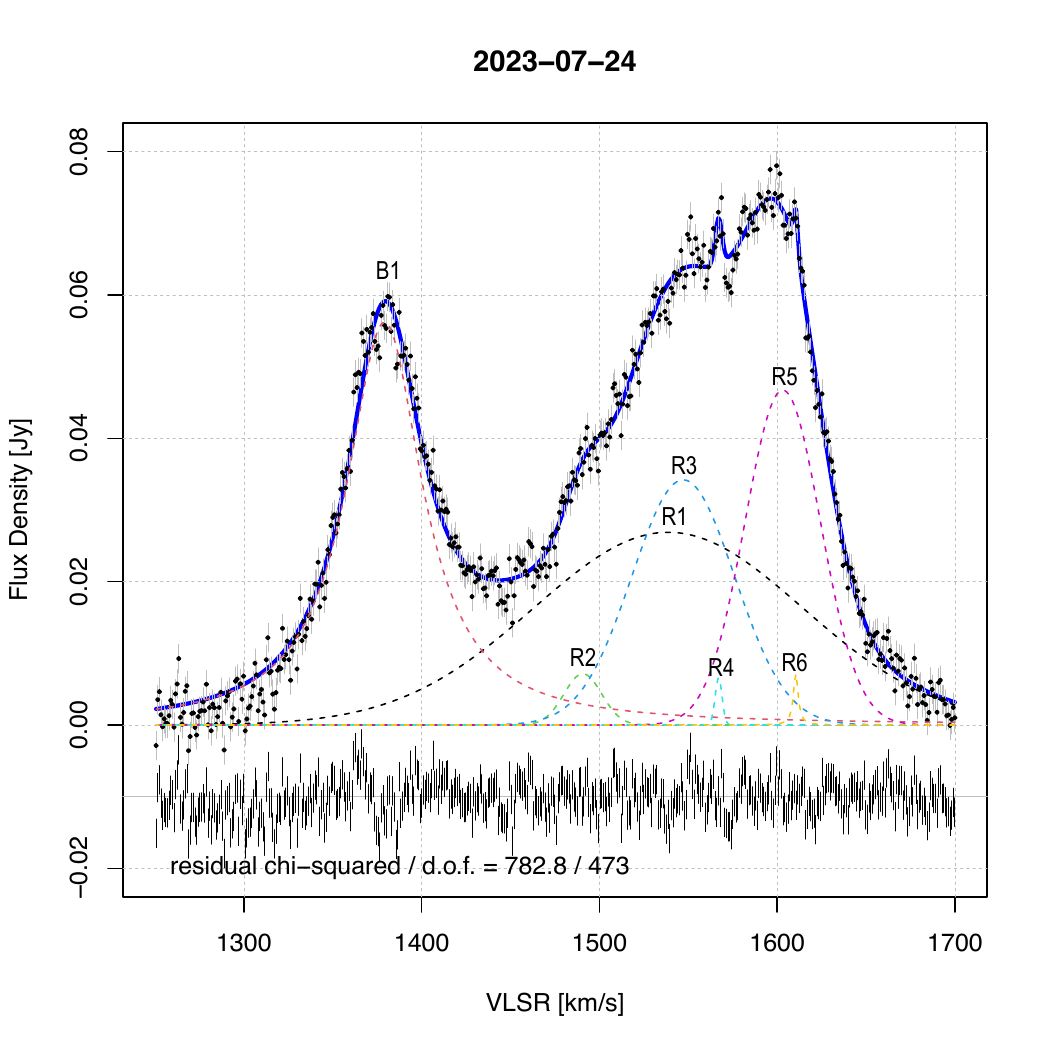}
 \end{center}
 \caption{Decomposition of the line profiles on 2023-07-10 (left) and 2023-07-24 (right). Solid blue lines show the best-fit model to the observed spectra shown in black filled circles with error bars. Dashed lines with labels represent decomposed components whose parameters are listed in Table \ref{tab:velocityComp}. The residuals are shown offset by $-0.01$ Jy.
 }\label{fig:velocityComponents}
\end{figure*}

To characterize the line profiles, we used the non-linear least squares function, {\tt nls}, in the R statistical language \citep{2023R} for spectral decomposition with 7 velocity components with either Gaussian or Lorentzian profile as summarized in table \ref{tab:velocityComp} and shown in figure \ref{fig:velocityComponents}.
We chose Gaussian or Lorentzian profile for each component to minimize the residuals.
Standard errors in table \ref{tab:velocityComp} were estimated by {\tt nls}.
See the source code {\tt plotMaserSpec.R} in the repository for detail.

The decompositions are not unique, especially in the red component consisting of complex sub-components.
Thus, the component labels in table \ref{tab:velocityComp} do not claim the unique identity between two epochs.
Residuals of the fittings results are characterized by $\chi^2$ over the degree of freedoms (d.o.f.) as shown in figure \ref{fig:velocityComponents}.

%-------- Observation Log Table
\begin{table*}
  \tbl{Spectral decomposition.}{%
  \begin{tabular}{lrrrcrrrl}
      \hline
      Component & \multicolumn{3}{c}{2023-07-10}                     && \multicolumn{3}{c}{2023-07-24}    & Type\footnotemark[$*$] \\ \cline{2-4} \cline{6-8}
	            & $V_{\rm center}$ & Peak          & FWHM            && $V_{\rm center}$ & Peak           & FWHM             &   \\
				& (km s$^{-1}$)    & (mJy)         & (km s$^{-1}$)   && (km s$^{-1}$)    & (mJy)          &  (km s$^{-1}$)   &   \\ \hline
	  B1       & $1374.2 \pm 0.4$ & $41.7 \pm 0.7$ & $ 50.9 \pm 1.3$ && $1379.1 \pm 0.3$ & $56.2 \pm 1.1$ & $ 52.1 \pm  1.2$ & L \\
	  R1       & $1530.1 \pm 2.0$ & $42.3 \pm 1.1$ & $113.9 \pm 3.7$ && $1538.7 \pm 4.8$ & $26.9 \pm 4.8$ & $178.4 \pm 15.8$ & G \\
	  R2       & $1485.6 \pm 1.0$ & $ 7.1 \pm 1.4$ & $ 10.9 \pm 2.6$ && $1490.7 \pm 1.7$ & $ 7.1 \pm 1.6$ & $ 25.0 \pm  5.3$ & G \\
	  R3       & $1525.3 \pm 0.7$ & $11.4 \pm 1.4$ & $ 11.9 \pm 1.9$ && $1547.1 \pm 3.0$ & $34.2 \pm 4.8$ & $ 68.1 \pm  8.0$ & G \\
	  R4       & $1552.1 \pm 0.6$ & $ 9.8 \pm 1.6$ & $  8.4 \pm 1.7$ && $1567.0 \pm 0.6$ & $ 6.6 \pm 1.7$ & $  4.6 \pm  1.5$ & G \\
	  R5       & $1595.6 \pm 0.6$ & $ 8.3 \pm 1.8$ & $  6.4 \pm 1.8$ && $1603.1 \pm 1.6$ & $46.7 \pm 3.8$ & $ 50.0 \pm  2.4$ & G \\
	  R6       & $1607.8 \pm 1.7$ & $27.1 \pm 2.1$ & $ 61.3 \pm 3.1$ && $1610.4 \pm 0.5$ & $ 7.2 \pm 2.2$ & $  3.2 \pm  1.7$ & L \\
      \hline
    \end{tabular}}\label{tab:velocityComp}
\begin{tabnote}
\footnotemark[$*$]G and L stand for Gaussian and Lorentzian functions, respectively.\\ 
\end{tabnote}
\end{table*}

Identification of the blue component B1 was certain as it was was isolated from others.
While the velocity width did not change significantly, the peak flux density increased by $34.5 \pm 4.2$\% and the center velocity shifted by $4.85 \pm 0.5$ km s$^{-1}$ in 14 days.

The red component was too complicated to identify every sub component in two epochs.
Four spiky sub components with the FWHM $<12$ km s$^{-1}$ (R2, R3, R4, and R5) on 2023-07-10 became unclear behind bumps composed by R3 and R5 on 2023-07-24.
We evaluated integrated intensity of the red components by subtracting the Lorentzian blue component model from the observed line profile.
Integrated flux densities of the red components increased from $7.69 \pm 0.23$ Jy km s$^{-1}$ to $10.12 \pm 0.30$ Jy km s$^{-1}$ by $31.6 \pm 4.2$\%.
Weighted mean velocity of the red components shifted from $1552.7 \pm 1.1$ km s$^{-1}$ to $1559.0 \pm 0.8$ km s$^{-1}$ by $6.3 \pm 1.4$ km s$^{-1}$.

\subsection{Size of emission regions}
\begin{figure*}
 \begin{center}
  \includegraphics[width=0.45\linewidth]{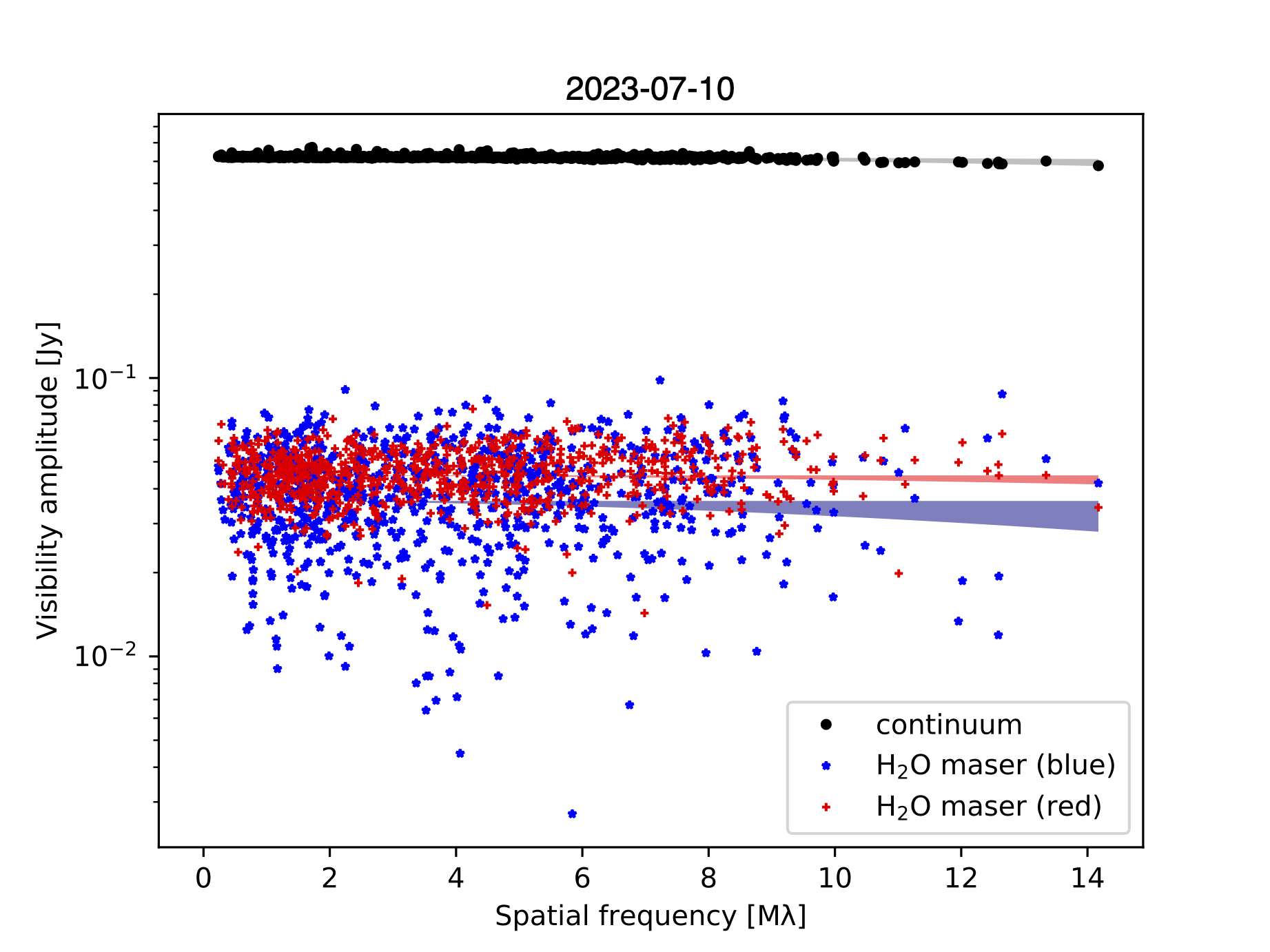}
  \includegraphics[width=0.45\linewidth]{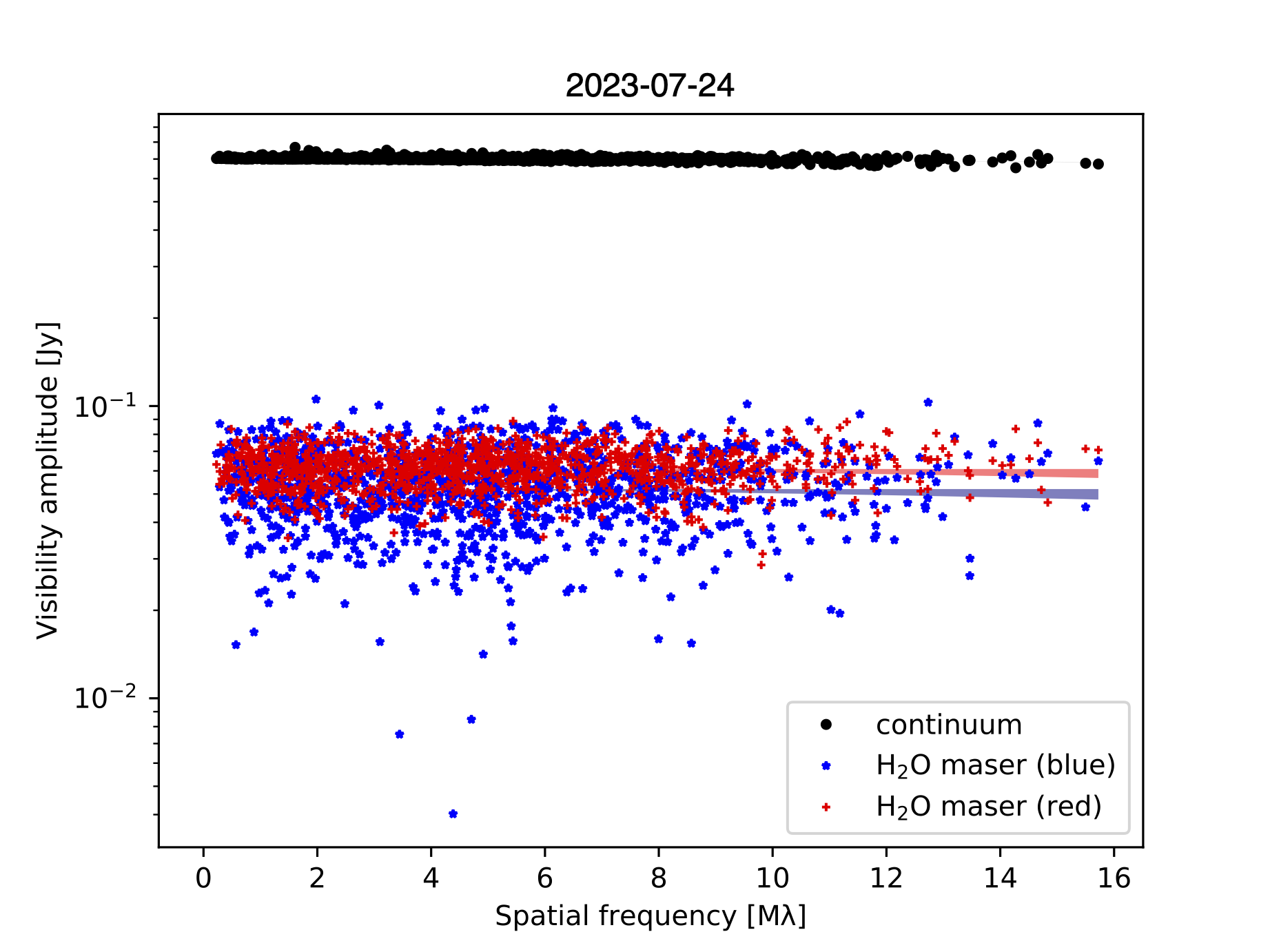}
 \end{center}
 \caption{Visibility amplitudes as a function of spatial frequency (projected baseline length / wavelength). The vertical axis is presented in logarithmic scale. Black, blue, and red markers stand for the continuum and H$_2$O line emission in velocity ranges of 1360 -- 1500 km s$^{-1}$ and 1500 -- 1625 km s$^{-1}$, respectively. The confidence bands indicate the best-fits and 99\% confidence levels.} \label{fig:uvplot}
\end{figure*}

We measured sizes of the continuum and line-emitting components by fitting a Gaussian function to visibilities as a function of spatial frequency (also known as $u,v$ distance).
Emission-line visibilities of blue and red components were generated by averaging the spectrum in the velocity ranges of 1360 -- 1500 km s$^{-1}$ and 1500 -- 1625 km s$^{-1}$, respectively.
Figure \ref{fig:uvplot} shows the visibility amplitude as a function of spatial frequencies, together with the confidence bands of the best fits and 99\% confidence levels.

The results are summarized in table \ref{tab:componentSize} together with the minimum brightness temperatures.
Broader spatial frequency coverage on 2023-07-24 allowed us to constrain the size of emission region in $<0.6$ milliarcsec (mas) and the brightness temperature of $T_{\rm B} > 10^6$ K.

%-------- Observation Log Table
\begin{table*}
  \tbl{Estimated Gaussian models and brightness temperatures of the continuum and line-emitting components}{%
  \begin{tabular}{lrrrrcrrrr}
      \hline
      Component & \multicolumn{4}{c}{2023-07-10}                               && \multicolumn{4}{c}{2023-07-24}         \\ \cline{2-5} \cline{7-10}
	            & Flux density  & Size (best)& Size (max) & $T_{\rm B}$               && Flux density  & Size (best)& Size (max) & $T_{\rm B}$ \\
				& (Jy)          & (mas)      & (mas)      & (K)                && (Jy)          & (mas)      & (mas)      & (K)  \\ \hline
	  Continuum & 0.624         & 0.51       & 0.65       & $>1.9 \times 10^7$ && 0.704         & 0.37       & 0.38       & $>6.2 \times 10^7$ \\
	  Blue      & 0.036         & 0.63       & 1.17       & $>3.1 \times 10^5$ && 0.052         & 0.35       & 0.60       & $>1.7 \times 10^6$\\
	  Red       & 0.045         & 0.30       & 0.63       & $>1.3 \times 10^6$ && 0.061         & 0.36       & 0.55       & $>2.4 \times 10^6$\\
      \hline
    \end{tabular}}\label{tab:componentSize}
\begin{tabnote}
A size stands for FWHM of a Gaussian component in milliarcsec (mas). Best and max sizes are derived by the most likelihood and 99\% confidence band shown in figure \ref{fig:uvplot}.

\end{tabnote}
\end{table*}

\subsection{Continuum images}
The continuum image on 2023-07-24 is shown in figure \ref{fig:ContMaserMap}. The image on 2023-07-10 is presented in figure \ref{fig:ContMaserMapJul10} in Appendix.
The flux densities, estimated from visibility amplitudes (figure \ref{fig:uvplot} and table \ref{tab:componentSize}), increased from $0.624 \pm 0.019$ Jy to $0.704 \pm 0.021$ Jy by $12.8 \pm 4.2$\%.
They were dominated by the unresolved core component with the peak intensities of 0.581 Jy and 0.649 Jy in the maps (figures \ref{fig:ContMaserMap} and \ref{fig:ContMaserMapJul10}).
The extended structure (`jets') in east (PA$=68^{\circ}.2$) and west (PA$=-111^{\circ}.5$) spans $\sim 0^{\prime \prime}.006$ to the east (PA$=68^{\circ}.2$) and west (PA$=-111^{\circ}.5$).
The extension is comparable to the size of the synthesized beam and is distinct in the super-resolution images.
The orientations were consistent with $61^{\circ}$ to $72^{\circ}.8$ and $-111^{\circ}.9$ to $-120^{\circ}.9$ for east and west jets in VLBI images \citep{2001PASJ...53..169K, 2003PASA...20..134K, 2003A&A...401..113V, 2004A&A...426..481K, 2019A&A...623A..27B, 2022A&A...658A.119B}.

\begin{figure*}
 \begin{center}
  \includegraphics[width=0.75\linewidth]{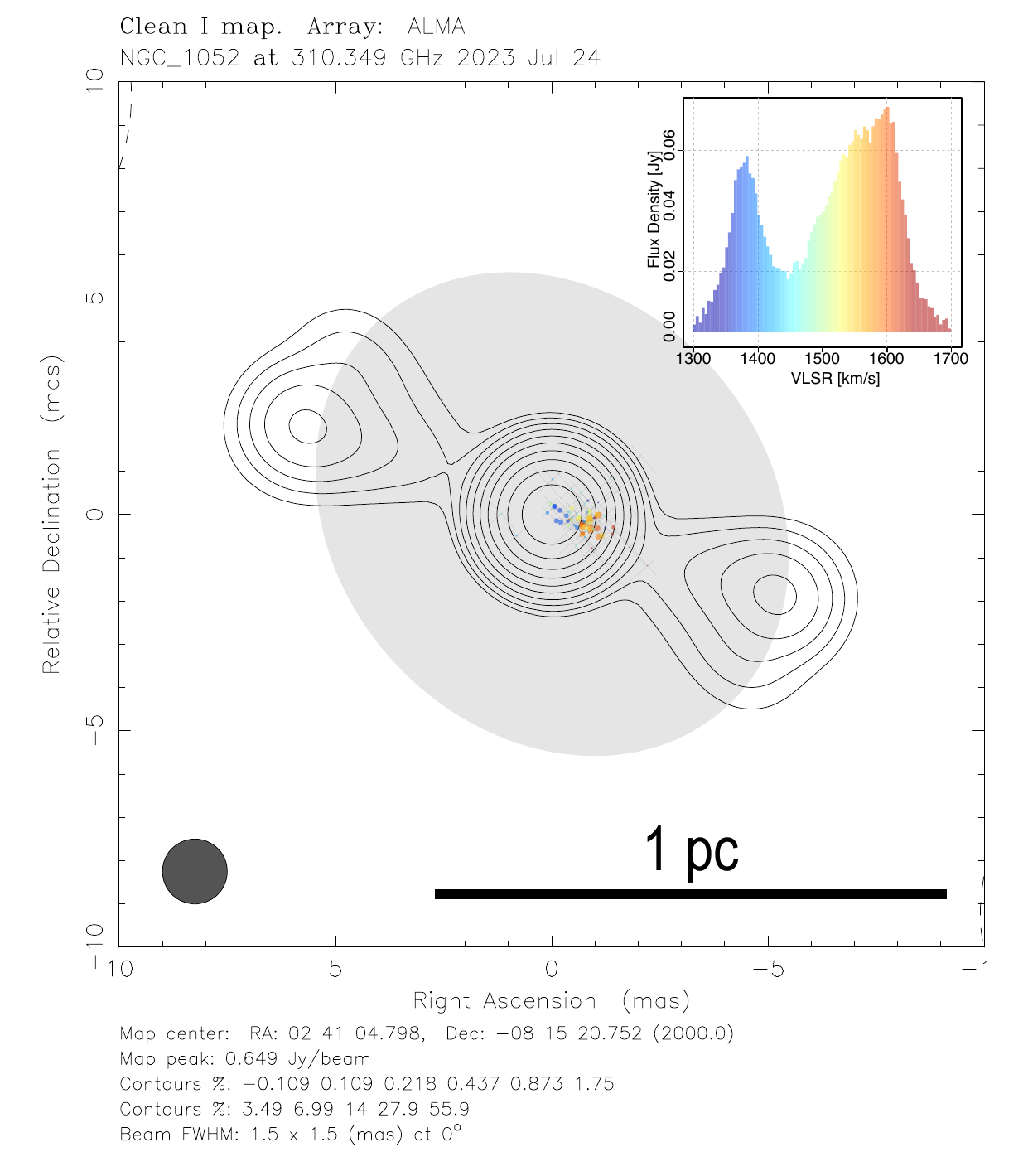}
 \end{center}
 \caption{Continuum and maser maps on 2023-07-24.
 The contours show continuum image at 310 GHz combining BBs 1--3.
 The contour levels are $\pm 3\sigma_{\rm cont}$ multiplied by powers of 2 where $\sigma_{\rm cont} = 0.24$ mJy beam$^{-1}$.
 The synthesized beam and the restoring beam are shown as the gray ellipse at the map center and the circle at the left bottom corner.
 Locations of maser components are drawn in filled colored circles with error bars. The marker size is proportional to line flux density and the color represents the velocity as shown in the spectral profiles in the upper right corner. The map on 2023-07-10 is presented in figure \ref{fig:ContMaserMapJul10} in Appendix.}\label{fig:ContMaserMap}
\end{figure*}

\subsection{Positioning line emitting components} \label{subsec:linePosition}
We performed {\tt uvmodelfit} task in CASA to estimate the positions of line emission source using continuum-subtracted visibilities binned by 5 spectral channels (4.57 km s$^{-1}$).
We set an unresolved point source model with a fixed flux density. Threshold for the maser flux density was set to $3\times$ the line image rms listed in table \ref{tab:obslog}.
Because self-calibration was applied before continuum subtraction, {\tt uvmodelfit} determined only relative positions of the line emitters with respect to the continuum core.

Figure \ref{fig:ContMaserMap} shows distribution of the emission line components on 2023-07-24 registered on the continuum map together with the binned spectrum as a color indicator of velocities. The same plot on 2023-07-10 is presented in figure \ref{fig:ContMaserMapJul10} in Appendix.
Close-up pictures are shown in figures \ref{fig:maserMap} and \ref{fig:ContMaserMapJul10}.
Median and interquartile ranges for position errors along major and minor axes were 0.43 (0.33 - 0.84) mas and 0.32 (0.24 - 0.63) mas on 2023-07-10, 0.23 (0.15 - 0.41) mas and 0.16 (0.11 - 0.29) mas on 2023-07-24, respectively.
Positions on 2023-07-10 were more scattered due to larger position errors compared with those on 2023-07-24.

\begin{figure*}
 \begin{center}
  \includegraphics[width=0.45\linewidth]{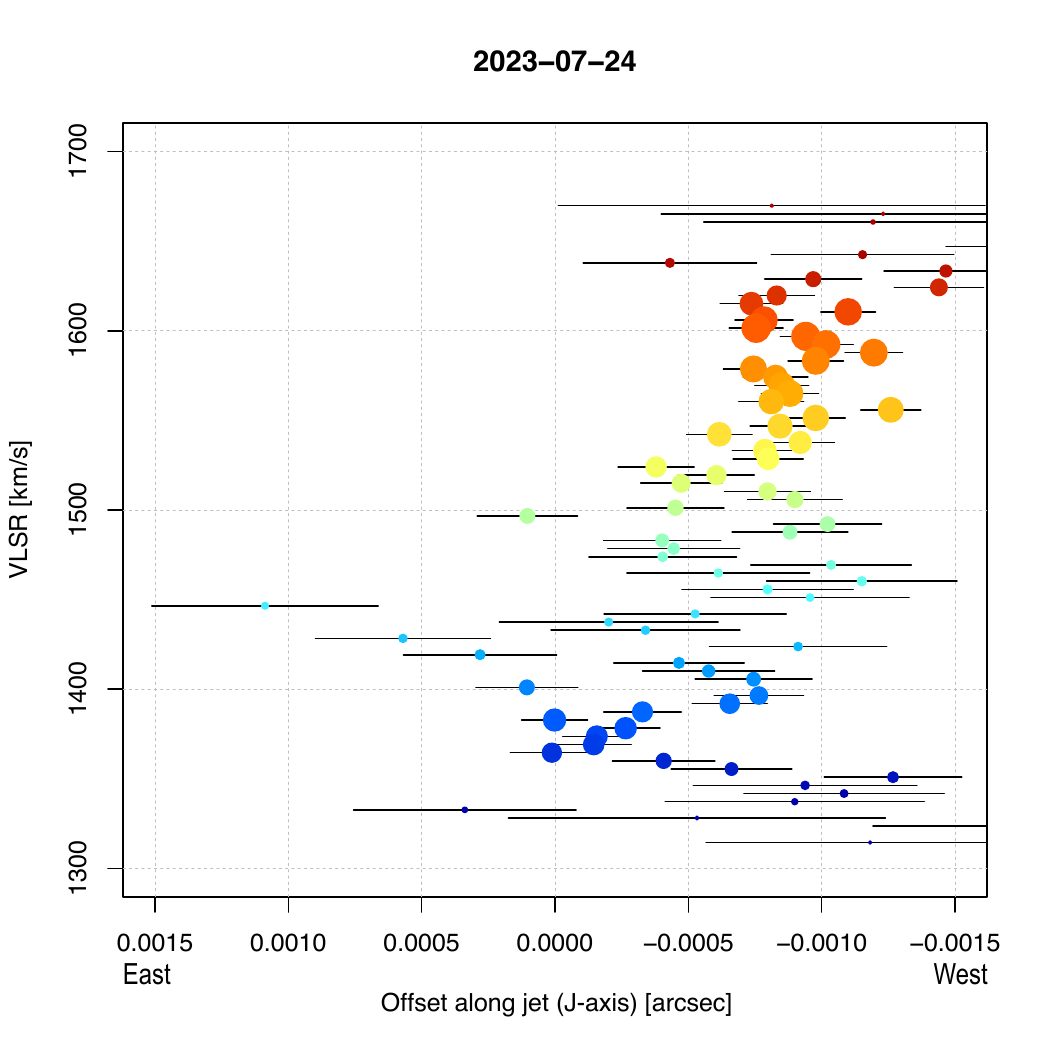}
  \includegraphics[width=0.45\linewidth]{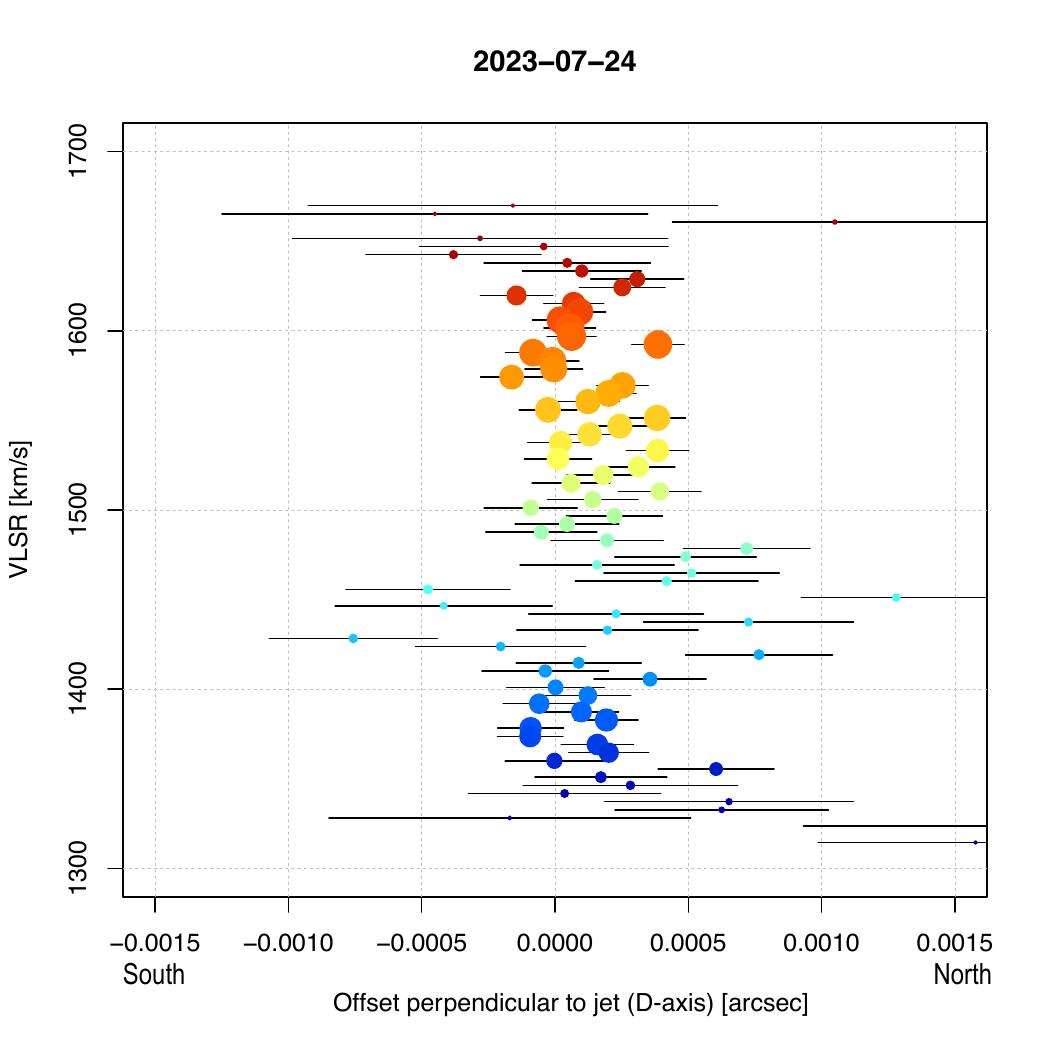}
 \end{center}
 \caption{Position--velocity diagrams along the jet (left, $J$-axis in Table \ref{tab:velocitygradient}) and perpendicular to the jet (right, $D$-axis in Table \ref{tab:velocitygradient}). The diagram on 2023-07-10 is presented in figure \ref{fig:PVdiagramJul10} in Appendix. Color and size of markers represents the velocity and the flux density as well as Fig.\ref{fig:ContMaserMap}.}\label{fig:PVdiagram}
\end{figure*}

We identified significant positional differences between blue and red components.
While blue components were centered at the continuum peak position, the center of red components offsets along the receding side of the jet by $\sim 1$ mas.
We applied linear regression of $V_{\rm LSR} \sim j J + d D$, where $J$ and $D$ are offset along PA=$68^{\circ}.35$ and $-21^{\circ}.65$, respectively, to estimate the velocity gradients, $j$ and $d$ along and perpendicular to the jet.
Figure \ref{fig:PVdiagram} shows position--velocity diagram on 2023-07-24 along ($J$-axis) and perpendicular ($D$-axis) to the jet. That on 2023-07-10 is presented in figure \ref{fig:PVdiagramJul10} in Appendix.
Best-fit results are summarized in table \ref{tab:velocitygradient}.
In both epochs, velocity gradient along the jet is significant while no significant gradient perpendicular to the jet was identified.
The sign of velocity gradient along the jet remains the same, though the values are significantly different between two epochs.
The smaller gradient in the first epoch can be caused by the larger positional scatter caused by poorer $(u, v)$ coverage.

Comparing with the spatial distribution of the 22-GHz H$_2$O maser consisting of two clusters separated by a 1-mas gap \citep{2008ApJ...680..191S}, the span of the 321-GHz maser fits the size of the 22-GHz gap though the absolute position is uncertain to register these maps.

\begin{figure*}
 \begin{center}
  \includegraphics[width=0.75\linewidth]{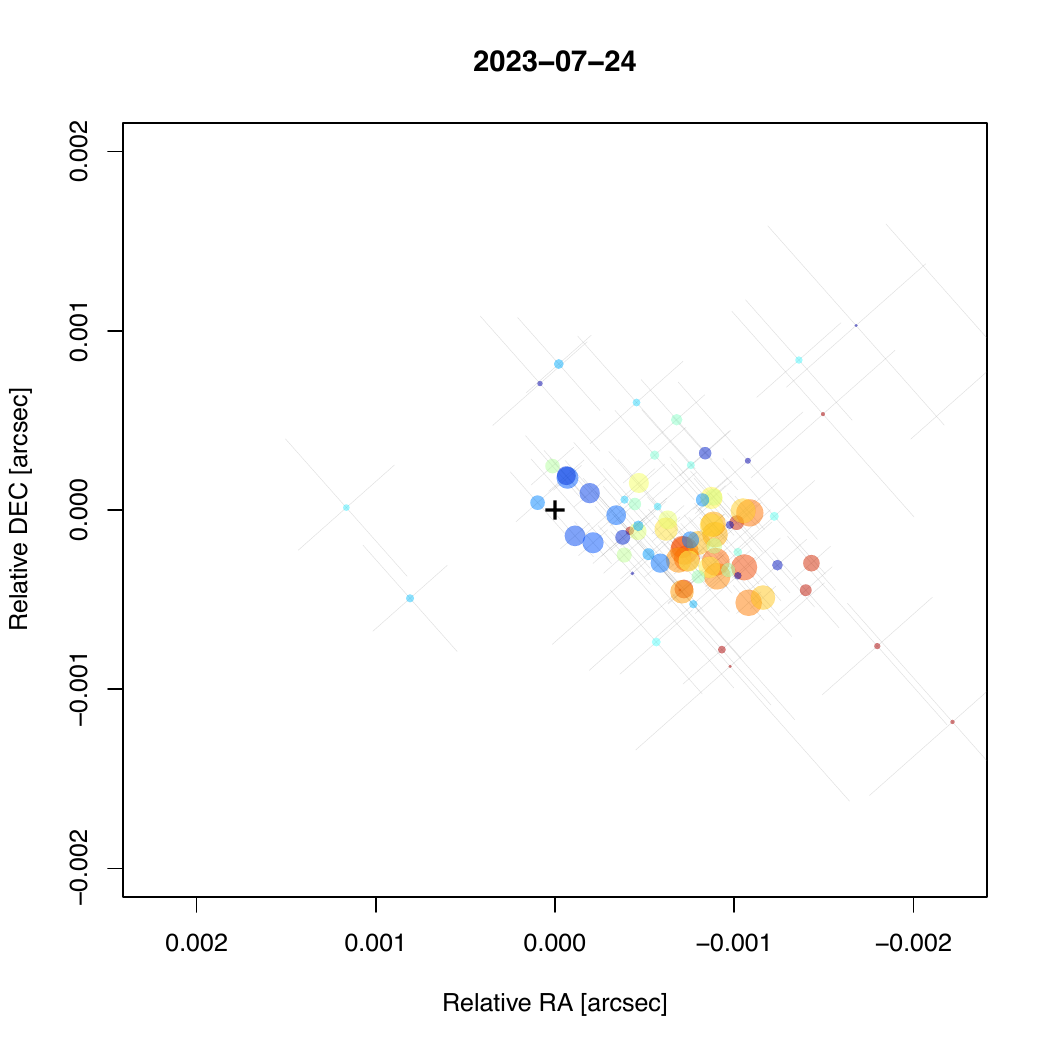}
 \end{center}
 \caption{Close-up of maser components on 2023-07-24. Coordinates are the same with figure \ref{fig:ContMaserMap}. The cross at the origin indicates the continuum peak position. The color represents the velocity as shown in the spectral profiles in the bottom left corner. The size of markers is proportional to flux density of at the channel. The plot on 2023-07-10 is presented in figure \ref{fig:ContMaserMapJul10} in Appendix. Color indicator is shown in figure \ref{fig:ContMaserMap}.}\label{fig:maserMap}
\end{figure*}

\begin{table*}
  \tbl{Linear regression for velocity gradient}{%
  \begin{tabular}{lrrcrr}
      \hline
      Epoch     & \multicolumn{2}{c}{2023-07-10}      && \multicolumn{2}{c}{2023-07-24}        \\ \cline{2-3} \cline{5-6}
	            & gradient      & $P$ value           && gradient      & $P$ value             \\
	  Axis	    & (km s$^{-1}$ mas$^{-1}$) &          && (km s$^{-1}$ mas$^{-1}$) &            \\ \hline
	  $J$       & $j = -94 \pm 19$  & $5.2\times 10^{-6}$ && $j = -173 \pm 19$  & $1.0\times 10^{-13}$ \\
	  $D$       & $d = 16 \pm 15$  & $0.29$               && $d = -18 \pm 36$  & $0.602$ \\
      \hline
    \end{tabular}}\label{tab:velocitygradient}
\begin{tabnote}
\end{tabnote}
\end{table*}

\section{Discussion}\label{sec:discussion}
\subsection{Justification for maser emission} \label{subsec:maseremission}
The ALMA long-baseline observations revealed the brightness temperature exceeding the excitation energy level and clearly rule out the possibility of thermal emission.
Strong intensity variation indicates that the emission is a maser.
The timescale of $\Delta T = 14$ days constrains the size of the emitter $< c\Delta T = 0.012$ pc.
The most plausible explanation for the compact, high-brightness non-thermal emission line on the continuum core, is maser amplification of the background continuum emission via population-inverted molecular gas.

\subsection{Velocity drift} \label{subsec:acceleration}
B1 maser component showed the velocity drift of $127 \pm 13$ km s$^{-1}$ yr$^{-1}$ or acceleration of $g = 4.0 \times 10^{-3}$ m s$^{-2}$.
The drift rate is an order of magnitude greater than that of well-known 22-GHz H$_2$O maser in NGC 4258; 7.5 km s$^{-1}$ yr$^{-1}$ \citep{1994ApJ...437L..35H}, $9.5 \pm 1.1$ km s$^{-1}$ yr$^{-1}$ \citep{1995A&A...304...21G}, or $9.6 \pm 1.0$ km s$^{-1}$ yr$^{-1}$ \citep{1995PASJ...47..771N}.

If B1 is identified as the 1333 km s$^{-1}$ component \citep{2023PASJ...75L...1K} on 2022-05-12, the velocity drift would be $35.3 \pm 0.9$ km s$^{-1}$ yr$^{-1}$.
Since the drift rate in 424 days is significantly smaller than that in 14 days, it is unlikely to identify the 1333 km s$^{-1}$ component as B1.

The velocity drift in red components is more complicated. The change of weighted mean velocity corresponds to $164 \pm 37$ km s$^{-1}$ yr$^{-1}$ or $5.2 \times 10^{-3}$ m s$^{-2}$.
The change is ascribed to either velocity drift or growth of R3 and R5 components.
Comparing with the 1467 km s$^{-1}$ component \citep{2023PASJ...75L...1K} on 2022-05-12, we have $73.8 \pm 4.4$ km s$^{-1}$ yr$^{-1}$ in 424 days, which is again inconsistent with the 14-day change.
Spectral monitoring with a higher cadence than 14 days is desired to trace the velocity drift of every component.

Nevertheless, all of velocity changes show the same direction, increase of radial velocity, are similar to the velocity drift of the systemic-velocity component of the 22-GHz H$_2$O maser disk in NGC 4258 \citep{1995A&A...304...21G,1995PASJ...47..771N}, indicating gravitational acceleration toward the nucleus as the masers locate in front of the core.

We then discuss about the velocity drift of B1.
If the velocity drift is ascribed to gravitational acceleration of the SMBH that weighs $M_{\rm BH} = 1.5 \times 10^8$ M$_{\Sol}$ \citep{2002ApJ...579..530W}, we have the distance of the maser from the SMBH, $r$, as $r = \sqrt{GM_{\rm BH}/g} = 0.072$ pc or $5000 \ R_{\rm s}$ where $R_{\rm s} = 2GM_{\rm BH}/c^2$ is the Schwarzschild radius.
While the radial velocity of B1 is blushifted with respect to the systemic velocity, the relative velocity of 118 km s$^{-1}$ is expected to vanish within 1 yr due to the deceleration of $127 \pm 13$ km s$^{-1}$.
Indeed, the maser velocity does not exceed the escape velocity, $v_{\rm esc} = \sqrt{2GM_{\rm BH}/r} = 4.2\times 10^3$ km s$^{-1}$.

Compared with the inner boundary of $0.12$ pc or $3.3\times 10^4$ Schwarzschild radii for the 22-GHz H$_2$O maser disk in NGC 4258 \citep{2005ApJ...626..104M}, the submillimeter maser distance in NGC 1052, scaled by the Schwarzschild radius, is 6.6 times as close as that in NGC 4258.
The ratio of kinetic temperatures for 321-GHz to 22-GHz maser excitation conditions follows $T_k \propto r^{-1/2}$, as 2000 K : 800 K $\sim \sqrt{6.6}:1$
As we expected in section \ref{sec:introduction}, the submillimeter H$_2$O maser is likely to arise at a closer distance to the central engine.

Alternatively, we admit a chance of misidentifying appearance and disappearance of independent maser components as the velocity drift.
To ensure identification of maser components, high-cadence monitoring for the maser spectrum is desired. 

\subsection{Velocity gradient} \label{subsec:velocitygradient}
As presented in section \ref{subsec:linePosition}, we identified velocity gradient along the jet.
While the blue components reside on the continuum core, the red components offset by $\sim 0.1$ pc.
The direction of the velocity gradient is consistent with the 100-pc-scale bipolar outflow traced by [O\emissiontype{III}] \citep{2005ApJ...629..131S,2022A&A...664A.135C} and by [Ar\emissiontype{II}] and [Ne\emissiontype{III}] \citep{2023arXiv230701252G}.
As argued in \citet{2023PASJ...75L...1K}, the velocity range of the submillimeter H$_2$O maser in NGC 1052 coincides with the submillimeter SO absorption lines at the estimated temperatures of $344 \pm 43$ K implying excitation by jet-torus interaction \citep{2023ApJ...944..156K}.
The observed velocity gradient supports this scenario.

The sub-pc-scale 22-GHz H$_2$O maser \citep{1998ApJ...500L.129C,2008ApJ...680..191S} shows opposite direction of the velocity gradient in the western cluster toward the receding jet, while the eastern cluster did not show obvious velocity gradient.
\citet{2008ApJ...680..191S} found that the 1-mas (0.1 pc) gap between two clusters of the 22-GHz H$_2$O maser coincides the plasma obscuring torus that caused FFA. 
As the FFA opacity is proportional to $\nu^{-2.1}$, where $\nu$ is the frequency, the opacity of FFA at 1 GHz, $\tau_{0} \simeq 1000$, yields $\tau \simeq 1.5$ at 22 GHz.
Thus, the 22-GHz H$_2$O emission inside the torus is obscured and the maser must locate outside the plasma torus where dynamical interaction with the jet does not significantly impact the radial velocity.
They claimed that the velocity gradient in the western cluster indicate acceleration infalling gas.
On the contrary at 321 GHz, the FFA opacity will be 0.005 and the submillimeter maser is visible at a 0.2-pc vicinity of the core inside the plasma where jet-torus interaction plays a role in the velocity field.
This explains the opposite directions of velocity gradient between 22-GHz and 321-GHz masers.
Gravitational acceleration by the central SMBH is the major factor in both inside and outside the torus to cause the velocity drift.

Alternative model ascribes the velocity gradient to projected velocity of a rotating disk, which successfully modeled for 22-GHz H$_2$O megamaser sources \citep{1995Natur.373..127M}.
However, we conclude that the velocity gradient does not represent such a rotating disk model by following reasons.
If the velocity drift of $g = 4.0 \times 10^{-3}$ m s$^{-2}$ is ascribed to gravitational acceleration and the derived distance is the radius of a rotating disk, we expect the rotating speed of $V_{\rm rot} = c \sqrt{R_{\rm s} / 2r} = 3000$ km s$^{-1}$ and the velocity gradient of $V_{\rm rot} / r = 4.2\times 10^4$ km s$^{-1}$ pc$^{-1}$ or $3.5\times 10^3$ km s$^{-1}$ mas$^{-1}$. The observed velocity gradient is not consistent with the expectation for the rotating disk model.
Furthermore, the orientation of the gradient along the jet is unlikely for a simple rotating disk model.

If there exists a rotating disk perpendicular to the jet, the width of the maser emission section to support the velocity width of $\Delta V \sim 400$ km s$^{-1}$ would be $r \Delta V / V_{\rm rot} = 0.01$ pc or 0.1 mas.
The position accuracy of the maser components in our observations were not sufficient to measure the expected velocity gradient in that width, as shown in figure \ref{fig:PVdiagram} (right).
VLBI observations up to 43 GHz \citep{2020AJ....159...14N, 2022A&A...658A.119B} measured the width of the jet and revealed that the jet is collimated in a cylindrical shape with a width of $1.3\times 10^3$ $R_s$ inside a distance of $10^4$ R$_s$ from the central engine.
The estimated maser emission section width of $\sim 700$ $R_s$ is comparable to the estimated jet width.
The constraint on the maser emission section supports the scenario that the population-inverted gas at the inner side of the torus amplifies the background continuum emission of the jet.
To trace the disk rotation, position accuracy of $\sim 0.01$ mas is desired.

A rotating disk with $V_{\rm rot} = 3000$ km s$^{-1}$ would yield high-velocity components at large impact parameters.
Expected frequency offsets of $\pm 3.2$ GHz are outside of the spectral coverage in our observations.
An expected impact parameter of $r = 0.072$ pc or 0.85 mas is too small to be resolved.
Future observations with the spectral setup targeting the terminal velocity component would clarify presence or absence of a rotating disk.
A proper motion of a rotating disk with $3000$ km s$^{-1}$ would be identified by 8000-km VLBI observations within 240/SNR days if position accuracy would be obtained by a synthesized beam size divided by a SNR.

\subsection{Jet-torus interaction} \label{subsec:jetmaser}
Following the former section, the velocity gradient along the jet axis is vindication of jet-torus interaction.
Here we argue about physical parameters following discussion in \citet{2023ApJ...944..156K}.
Adopting overpressure factor of 1.5 \citep{2019A&A...629A...4F}, we have $\rho_j v^2_j = 1.5 \rho_t v^2_s$, where $\rho_j$ and $v_j$ are density and speed of the jet, and $\rho_t$ and $v_s$ are the density of the torus and the advancing speed of the shock, respectively.
Taking $v_j = 0.26 c$ \citep{2003A&A...401..113V}, $\rho_t = 1.1 \times 10^{-12}$ kg m$^{-3}$ estimated from the maser excitation condition of $n_{\rm H2} \gtrsim 3.3 \times 10^8$ cm$^{-3}$ \citep{2016MNRAS.456..374G}, we have $\rho_j = 3.5 \times 10^{-18}$ kg m$^{-3}$.
This is one order of magnitude greater than the estimation by SO absorption feature \citep{2023ApJ...944..156K}.
Since the required temperature for the submillimeter H$_2$O maser is significantly higher than $344 \pm 43$ K estimated by submillimeter SO absorption lines, the population-inversion zone is likely to locate upstream of the jet compared with the SO evaporation region.
Higher pressure in the population-inversion zone than that in the SO evaporation region may indicate collimation and acceleration of the jet via jet-torus interaction.

\begin{figure*}
 \begin{center}
  \includegraphics[width=\linewidth]{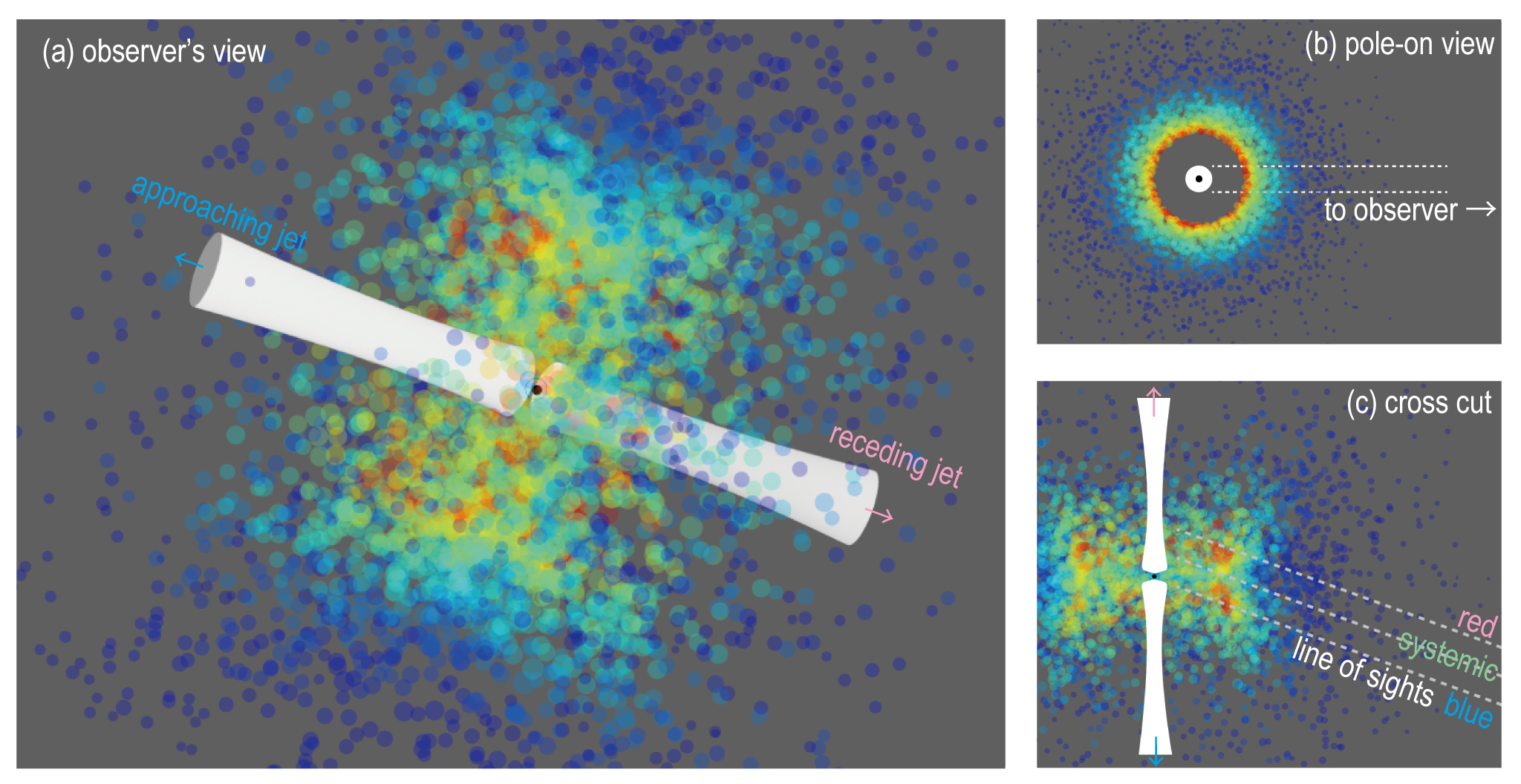}
 \end{center}
 \caption{A schematic of the clumpy torus and jets in NGC 1052. The black marker at the center indicates the core which emanates double-sided jets presented in white. The torus surrounding the core consists of molecular clumps (filled circles) and plasma (unseen filler). Color of clumps indicates excitation condition: population-inversion zone in red to orange and thermal gas in yellow, green, and blue. While yellow clumps reside in warm zone, green and blue clumps in the equatorial plane and outer area of the torus are colder. The population-inversion zone amplifies the background continuum emission (core and the jets) to produce the 321-GHz H$_2$O maser. The thermal gas casts molecular absorption lines on the background emission.
 (a):observer's view with a viewing angle of $70^{\circ}$. (b):pole-on view of the torus. The dashed lines stand for the jet width and the maser amplification area in the cross section with the population-inversion zone. (c):cross cut of the torus and the jet. Lines of sight through the torus stand for blue and red maser components while systemic velocity component is attenuated through the thermal gas on the plane of the equator.}\label{fig:ClumpyTorus}
\end{figure*}

\subsection{Torus structure} \label{subsec:torusstructure}
This section addresses clumpy torus structure based on the time variation of the maser profile.

The blue and red components of H$_2$O maser showed amplitude increase by $34.5 \pm 4.2$\% and $31.6 \pm 4.2$\% in 14 days while the continuum flux density increased by $12.8 \pm 4.2$\%.
As maser intensity is a product of the brightness of the background emission source and the amplification (or absorption) gain through population-inversion (or thermal) zone, the breakdown of the increase of maser intensity involves $\sim 20$\% increase of the maser amplification gain variation with the timescale of $1.2\times 10^6$ s$ / \log 1.2 = 6.6 \times 10^{6}$ s.

The variation in amplification gain indicates inhomogeneity inside the molecular torus.
VLBI images for spatially resolved HCN \citep{2016ApJ...830L...3S} and HCO$^{+}$ \citep{2019ApJ...872L..21S} absorption lines estimated the upper limit of the clump size of $\le$ 0.1 pc in the molecular torus of NGC 1052. 
ALMA millimeter/submillimeter absorption-line studies \citep{2020ApJ...895...73K} showed that the molecular gas in the torus is clumpy with an estimated covering factor of $0.17^{+0.06}_{-0.03}$.
Passage of population-inverted clumps across the line of sight toward the continuum background can cause variation of amplification gain.
The Jeans length in the population-inversion zone is estimated to be $\lambda_J = c_s / \sqrt{G\rho} \sim 5.3 \times 10^{14}$ m or 0.017 pc, where $\rho = 1.1 \times 10^{-12}$ kg m$^{-3}$ is the density for maser excitation condition of $n_{\rm H2} > 3.3 \times 10^8$ cm$^{-3}$ \citep{2016MNRAS.456..374G} and $c_s$ is the sound speed for the kinematic temperature of 2000 K.
Crossing time of $\lambda_J / V_{\rm rot} = 1.8\times 10^8$ s is 27 times as long as the variation timescale.
Thus, passage of a population-inverted clump in front of the continuum background is unlikely to explain the maser gain variation.

Another possibility is the change of excitation condition supported by the increase of the continuum flux density.
Increase of depth in the population-inversion zone, generated by interaction with more powerful jet, can cause increase of the maser flux density.

Attenuation through thermal molecular gas outside the population-inversion zone would be another origin of the gain variation.
The spectral minimum near the systemic velocity can be caused by self absorption through thermal gas \citep{1994ApJ...432L..35W}.
\citet{2023ApJ...944..156K} estimated the temperatures of SO molecules showing millimeter and submillimeter absorption lines to be $26 \pm 4$ K and $344 \pm 43$ K, respectively.
Smaller Jeans length for lower-temperature yields shorter crossing time by one order of magnitude than that of a population-inverted clump and thus comparable to the maser variation timescale.

Figure \ref{fig:ClumpyTorus} illustrates the clumpy torus model.
Inside of the torus is excited by interaction with the jet to form population-inversion zone.
Outer region of the torus consists of colder thermal clumps that attenuate maser emission to bear the spectral minimum near the systemic velocity and absorption lines in thermal transitions.

\section{Summary}\label{sec:summary}
ALMA follow-up observations with the widest array configuration led findings summarized below.

\begin{enumerate}
\item The 321-GHz H$_2$O emission in NGC 1052 is definitely a maser with the brightness temperature $T_{\rm B} > 10^6$ K. The intensity variation in 14 days also supports that the emission is a maser. This indicates feasibility for VLBI observations to clarify spatially-resolved structures and proper motions of the maser together with the core and the jets.
A 8000-km-baseline array would offer a resolution of 0.002 pc which allows resolving the structure of maser emitting gas as discussed in the previous section.

\item The maser profile consists of mainly two velocity components, blue and red, straddling a local minimum near the systemic velocity. Both velocity components increased the flux density compared with the previous observation 424 days ago. They still grew by $34.5 \pm 4.2$\% and $31.6 \pm 4.2$\% in 14 days while the continuum flux density increased by $12.8 \pm 4.2$\%.
The variability in amplification gain implies inhomogeneity of the population-inversion zone and/or change of excitation condition.

\item Both velocity components showed redward velocity drifts. While the red component was too complex to distinguish the drift from growth of R3 and R5 sub-components, the velocity drift of $127 \pm 13$ km s$^{-1}$ yr$^{-1}$ in the blue component is more confident.
If the drift is real and is ascribed to gravitational acceleration onto the SMBH with $M_{\rm BH} = 1.5 \times 10^8$ M$_{\Sol}$, the distance of the maser will be 5000 $R_s$ from the SMBH.
Since the maser velocity relative to the systemic velocity is smaller than the escape velocity, the maser emitting gas must be bound in the SMBH gravitational potential.
For a rotating disk model, corresponding rotation velocity will be $V_{\rm rot} = 3000$ km s$^{-1}$.

\item We identified velocity gradient of $-174 \pm 19$ km s$^{-1}$ mas$^{-1}$ along the jet. The gradient indicates that the population-inverted gas is driven by the jet. No significant velocity gradient perpendicular to the jet was identified under positional errors of $\sim 0.16$ mas. To resolve expected velocity gradient for a rotating disk, a spatial resolution sharper than 0.1 mas is required.

\item The pressure of population-inversion zone is one order of magnitude greater than that of the SO evaporation region. This implies that the maser locates upstream of the jet compared with the SO evaporation region and jet acceleration and collimation process by jet-torus interaction.

\end{enumerate}

For future works, spectral monitoring is desired to ensure the velocity drift possibly caused by gravitational acceleration.
Designed spectral setup targeting the terminal velocity emission at $\sim \pm 3000$ km s$^{-1}$ would allow us to determine the rotation velocity, if the maser associates in a rotating disk.
VLBI observations are feasible to resolve the structure of the population-inverted zone and detect a proper motion for disk rotation or outflows.
Submillimeter masers would yield a potential to clarify the inner edge of molecular torus and to determine geometric distances.

\section*{Acknowledgments}
This paper makes use of the following ALMA data: ADS/JAO.ALMA\#2021.1.00341.S and ADS/JAO.ALMA\#2023.A.00023.S. ALMA is a partnership of ESO (representing its member states), NSF (USA) and NINS (Japan), together with NRC (Canada), MOST and ASIAA (Taiwan), and KASI (Republic of Korea), in cooperation with the Republic of Chile. The Joint ALMA Observatory is operated by ESO, AUI/NRAO and NAOJ.
This work is supported by JSPS KAKENHI 18K03712 and 21H01137.

\appendix
\section*{Results on 2023-07-10}
\begin{figure*}
 \begin{center}
  \includegraphics[width=0.4\linewidth]{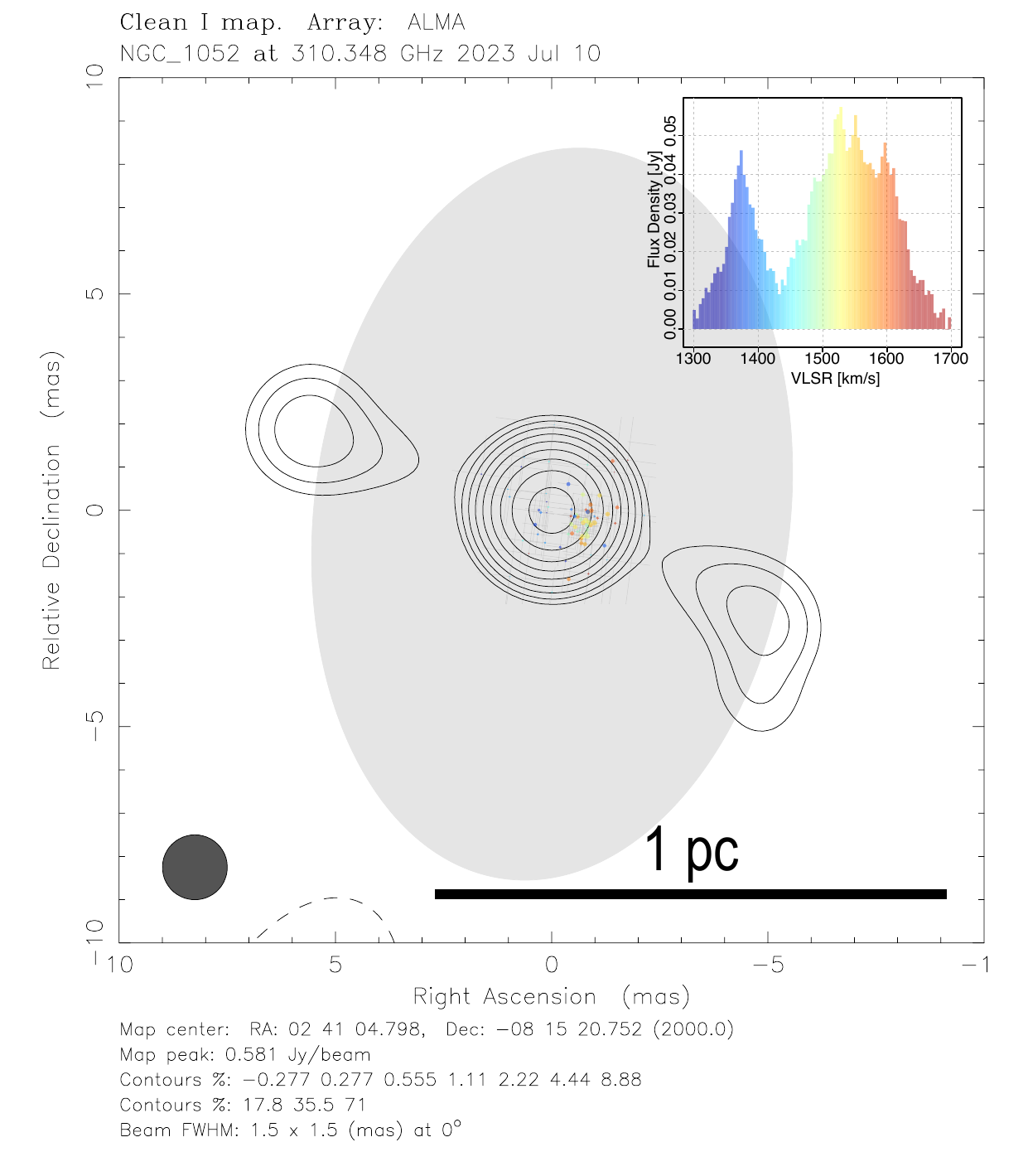}
  \includegraphics[width=0.5\linewidth]{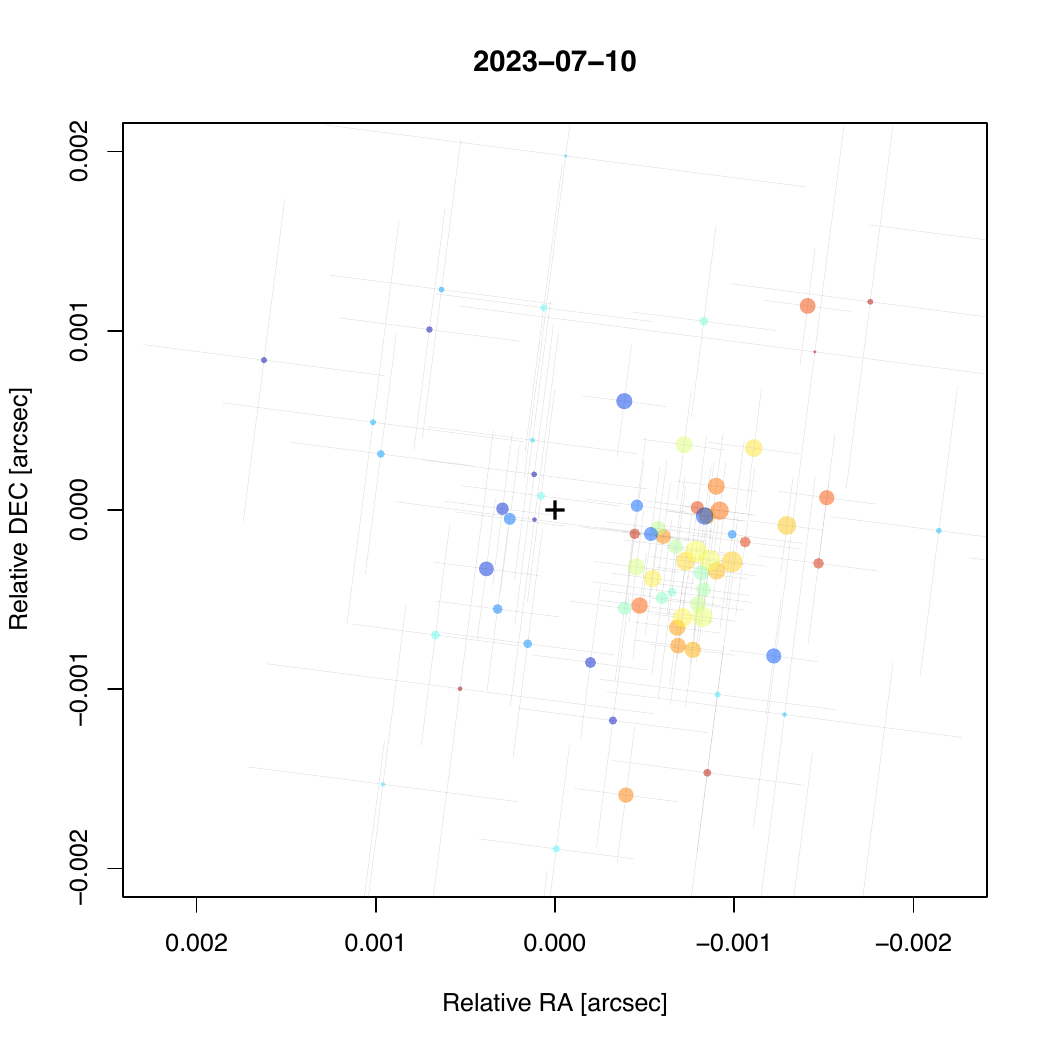}
 \end{center}
 \caption{Continuum and maser map (left) and close-up of maser components (right) on 2023-07-10, as well as figures \ref{fig:ContMaserMap} and \ref{fig:maserMap}. The contour levels are $\pm 3\sigma_{\rm cont}$ multiplied by powers of 2 where $\sigma_{\rm cont} = 0.57$ mJy beam$^{-1}$.}\label{fig:ContMaserMapJul10}
\end{figure*}

\begin{figure*}
 \begin{center}
  \includegraphics[width=0.45\linewidth]{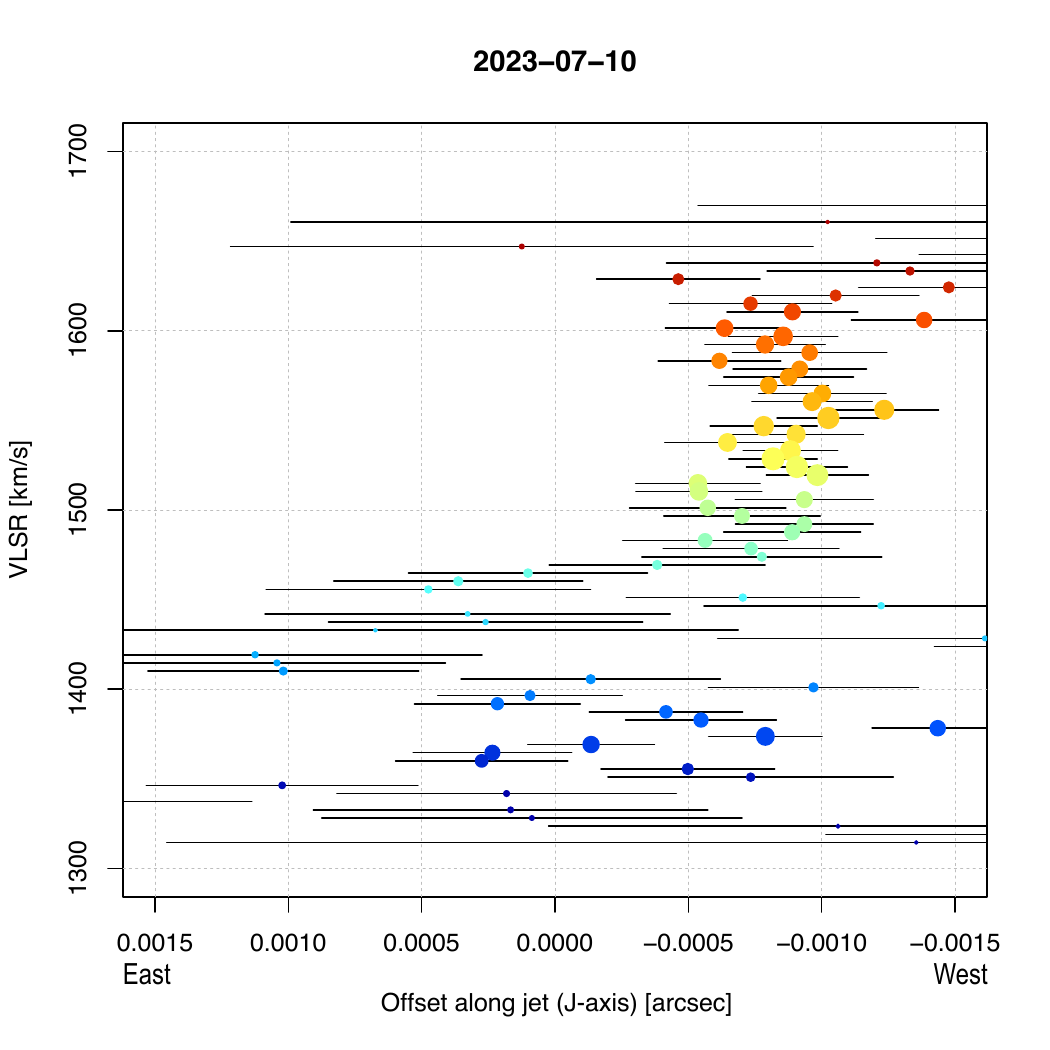}
  \includegraphics[width=0.45\linewidth]{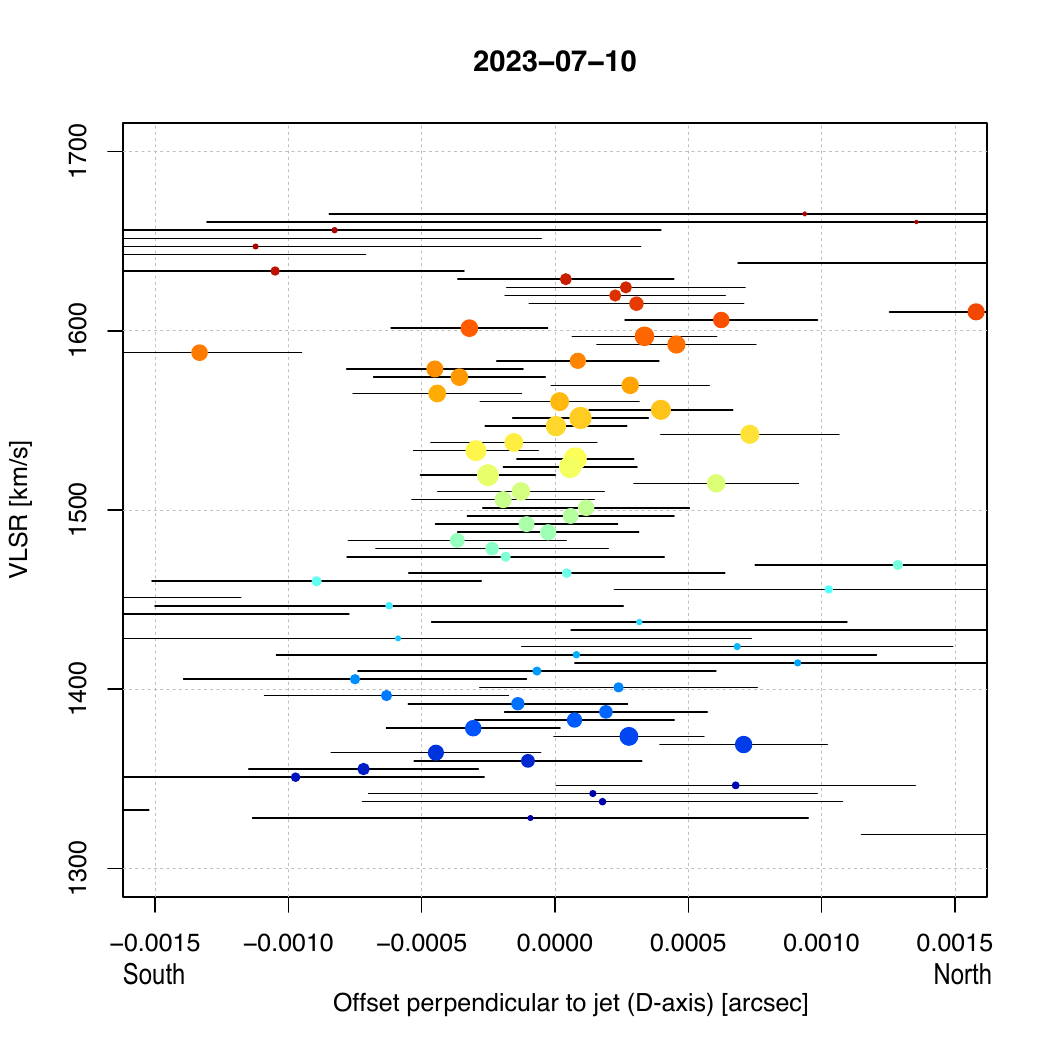}
 \end{center}
 \caption{Position--velocity diagrams along and perpendicular to the jet on 2023-07-10, as well as figure \ref{fig:PVdiagram}.}\label{fig:PVdiagramJul10}
\end{figure*}

The main text addresses results on 2023-07-24 for discussion about the maser distribution and the velocity gradient because the image quality is significantly better than that on 2023-07-10. They are presented in figures \ref{fig:ContMaserMapJul10} and \ref{fig:PVdiagramJul10}.

\begin{thebibliography}{}
% Journals(e.g. A\&A,ApJ,AJ,NMRAS,PASP ...)
% Authors, Year, Journal, Vol#, Page#
% Journal Title Abbreviation >> http://www.asj.or.jp/pasj/Jabb.html
%\bibitem[Baczko et al.(2016)]{2016A&A...593A..47B} Baczko, A.-K., Schulz, R., Kadler, M., et al.\ 2016, \aap, 593, A47%. doi:10.1051/0004-6361/201527951

\bibitem[Baczko et al.(2019)]{2019A&A...623A..27B} Baczko, A.-K., Schulz, R., Kadler, M., et al.\ 2019, \aap, 623, A27%. doi:10.1051/0004-6361/201833828

\bibitem[Baczko et al.(2022)]{2022A&A...658A.119B} Baczko, A.-K., Ros, E., Kadler, M., et al.\ 2022, \aap, 658, A119%. doi:10.1051/0004-6361/202141897

\bibitem[Braatz et al.(1994)]{1994ApJ...437L..99B} Braatz, J.~A., Wilson, A.~S., \& Henkel, C.\ 1994, \apjl, 437, L99%. doi:10.1086/187692

\bibitem[Braatz et al.(2003)]{2003ApJS..146..249B} Braatz, J.~A., Wilson, A.~S., Henkel, C., et al.\ 2003, \apjs, 146, 249%. doi:10.1086/374417

\bibitem[CASA Team et al.(2022)]{2022PASP..134k4501C} CASA Team, Bean, B., Bhatnagar, S., et al.\ 2022, \pasp, 134, 114501%. doi:10.1088/1538-3873/ac9642

\bibitem[Cazzoli et al.(2022)]{2022A&A...664A.135C} Cazzoli, S., Hermosa Mu{\~n}oz, L., M{\'a}rquez, I., et al.\ 2022, \aap, 664, A135%. doi:10.1051/0004-6361/202142695

\bibitem[Claussen et al.(1998)]{1998ApJ...500L.129C} Claussen, M.~J., Diamond, P.~J., Braatz, J.~A., et al.\ 1998, \apjl, 500, L129%. doi:10.1086/311405

\bibitem[Fromm et al.(2019)]{2019A&A...629A...4F} Fromm, C.~M., Younsi, Z., Baczko, A., et al.\ 2019, \aap, 629, A4%. doi:10.1051/0004-6361/201834724

\bibitem[Gallimore \& Impellizzeri(2023)]{2023ApJ...951..109G} Gallimore, J.~F. \& Impellizzeri, C.~M.~V.\ 2023, \apj, 951, 109%. doi:10.3847/1538-4357/acd846

\bibitem[Goold et al.(2023)]{2023arXiv230701252G} Goold, K., Seth, A., Molina, M., et al.\ 2023, arXiv:2307.01252%. doi:10.48550/arXiv.2307.01252

\bibitem[Gray et al.(2016)]{2016MNRAS.456..374G} Gray, M.~D., Baudry, A., Richards, A.~M.~S., et al.\ 2016, \mnras, 456, 374%. doi:10.1093/mnras/stv2437

\bibitem[Greenhill et al.(1995)]{1995A&A...304...21G} Greenhill, L.~J., Henkel, C., Becker, R., et al.\ 1995, \aap, 304, 21

\bibitem[Hagiwara et al.(2013)]{2013ApJ...768L..38H} Hagiwara, Y., Miyoshi, M., Doi, A., et al.\ 2013, \apjl, 768, L38%. doi:10.1088/2041-8205/768/2/L38

\bibitem[Hagiwara et al.(2016)]{2016ApJ...827...69H} Hagiwara, Y., Horiuchi, S., Doi, A., et al.\ 2016, \apj, 827, 69%. doi:10.3847/0004-637X/827/1/69

\bibitem[Hagiwara et al.(2021)]{2021ApJ...923..251H} Hagiwara, Y., Horiuchi, S., Imanishi, M., et al.\ 2021, \apj, 923, 251%. doi:10.3847/1538-4357/ac3089

\bibitem[Haschick et al.(1994)]{1994ApJ...437L..35H} Haschick, A.~D., Baan, W.~A., \& Peng, E.~W.\ 1994, \apjl, 437, L35%. doi:10.1086/187676

\bibitem[Herrnstein et al.(2005)]{2005ApJ...629..719H} Herrnstein, J.~R., Moran, J.~M., Greenhill, L.~J., et al.\ 2005, \apj, 629, 719%. doi:10.1086/431421

\bibitem[Humphreys et al.(2005)]{2005ApJ...634L.133H} Humphreys, E.~M.~L., Greenhill, L.~J., Reid, M.~J., et al.\ 2005, \apjl, 634, L133%. doi:10.1086/498890

\bibitem[Humphreys et al.(2008)]{2008ApJ...672..800H} Humphreys, E.~M.~L., Reid, M.~J., Greenhill, L.~J., et al.\ 2008, \apj, 672, 800%. doi:10.1086/523637

\bibitem[Humphreys et al.(2013)]{2013ApJ...775...13H} Humphreys, E.~M.~L., Reid, M.~J., Moran, J.~M., et al.\ 2013, \apj, 775, 13%. doi:10.1088/0004-637X/775/1/13

\bibitem[Impellizzeri(2022)]{2022NatAs...6..885I} Impellizzeri, C.~M.~V.\ 2022, Nature Astronomy, 6, 885%. doi:10.1038/s41550-022-01764-2

\bibitem[Impellizzeri et al.(2008)]{2008evn..confE..33I} Impellizzeri, V., Roy, A.~L., \& Henkel, C.\ 2008, The role of VLBI in the Golden Age for Radio Astronomy, 9, 33%. doi:10.22323/1.072.0033

\bibitem[Kadler et al.(2004)]{2004A&A...426..481K} Kadler, M., Ros, E., Lobanov, A.~P., et al.\ 2004, \aap, 426, 481%. doi:10.1051/0004-6361:20041051

\bibitem[Kameno et al.(2001)]{2001PASJ...53..169K} Kameno, S., Sawada-Satoh, S., Inoue, M., et al.\ 2001, \pasj, 53, 169%. doi:10.1093/pasj/53.2.169

\bibitem[Kameno et al.(2003)]{2003PASA...20..134K} Kameno, S., Inoue, M., Wajima, K., et al.\ 2003, \pasa, 20, 134%. doi:10.1071/AS03003

\bibitem[Kameno et al.(2005)]{2005ApJ...620..145K} Kameno, S., Nakai, N., Sawada-Satoh, S., et al.\ 2005, \apj, 620, 145%. doi:10.1086/426936

\bibitem[Kameno et al.(2020)]{2020ApJ...895...73K} Kameno, S., Sawada-Satoh, S., Impellizzeri, C.~M.~V., et al.\ 2020, \apj, 895, 73%. doi:10.3847/1538-4357/ab8bd6

\bibitem[Kameno et al.(2023a)]{2023ApJ...944..156K} Kameno, S., Sawada-Satoh, S., Impellizzeri, C.~M.~V., et al.\ 2023a, \apj, 944, 156%. doi:10.3847/1538-4357/acb499

\bibitem[Kameno et al.(2023b)]{2023PASJ...75L...1K} Kameno, S., Harikane, Y., Sawada-Satoh, S., et al.\ 2023b, \pasj, 75, L1%. doi:10.1093/pasj/psad011

\bibitem[Koekemoer et al.(1995)]{1995Natur.378..697K} Koekemoer, A.~M., Henkel, C., Greenhill, L.~J., et al.\ 1995, \nat, 378, 697%. doi:10.1038/378697a0

\bibitem[Liszt \& Lucas(2004)]{2004A&A...428..445L} Liszt, H. \& Lucas, R.\ 2004, \aap, 428, 445%. doi:10.1051/0004-6361:20041650

\bibitem[Miyoshi et al.(1995)]{1995Natur.373..127M} Miyoshi, M., Moran, J., Herrnstein, J., et al.\ 1995, \nat, 373, 127%. doi:10.1038/373127a0

\bibitem[Modjaz et al.(2005)]{2005ApJ...626..104M} Modjaz, M., Moran, J.~M., Kondratko, P.~T., et al.\ 2005, \apj, 626, 104%. doi:10.1086/429559

\bibitem[Moran et al.(1995)]{1995PNAS...9211427M} Moran, J., Greenhill, L., Herrnstein, J., et al.\ 1995, Proceedings of the National Academy of Science, 92, 11427%. doi:10.1073/pnas.92.25.11427

\bibitem[Nakai et al.(1993)]{1993Natur.361...45N} Nakai, N., Inoue, M., \& Miyoshi, M.\ 1993, \nat, 361, 45%. doi:10.1038/361045a0

\bibitem[Nakai et al.(1995)]{1995PASJ...47..771N} Nakai, N., Inoue, M., Miyazawa, K., et al.\ 1995, \pasj, 47, 771

\bibitem[Nakahara et al.(2020)]{2020AJ....159...14N} Nakahara, S., Doi, A., Murata, Y., et al.\ 2020, \aj, 159, 14%. doi:10.3847/1538-3881/ab465b

\bibitem[Neufeld \& Melnick(1991)]{1991ApJ...368..215N} Neufeld, D.~A. \& Melnick, G.~J.\ 1991, \apj, 368, 215%. doi:10.1086/169685

\bibitem[Omar et al.(2002)]{2002A&A...381L..29O} Omar, A., Anantharamaiah, K.~R., Rupen, M., et al.\ 2002, \aap, 381, L29%. doi:10.1051/0004-6361:20011604

\bibitem[Pesce et al.(2016)]{2016ApJ...827...68P} Pesce, D.~W., Braatz, J.~A., \& Impellizzeri, C.~M.~V.\ 2016, \apj, 827, 68%. doi:10.3847/0004-637X/827/1/68

\bibitem[R Core Team (2023)]{2023R} R Foundation for Statistical Computing, Vienna, Austria. https://www.R-project.org/

\bibitem[Sawada-Satoh et al.(2008)]{2008ApJ...680..191S} Sawada-Satoh, S., Kameno, S., Nakamura, K., et al.\ 2008, \apj, 680, 191%. doi:10.1086/587886

\bibitem[Sawada-Satoh et al.(2016)]{2016ApJ...830L...3S} Sawada-Satoh, S., Roh, D.-G., Oh, S.-J., et al.\ 2016, \apjl, 830, L3%. doi:10.3847/2041-8205/830/1/L3

\bibitem[Sawada-Satoh et al.(2019)]{2019ApJ...872L..21S} Sawada-Satoh, S., Byun, D.-Y., Lee, S.-S., et al.\ 2019, \apjl, 872%, L21. doi:10.3847/2041-8213/ab0425

\bibitem[Shepherd et al.(1994)]{1994BAAS...26..987S} Shepherd, M.~C., Pearson, T.~J., \& Taylor, G.~B.\ 1994, \baas

\bibitem[Shostak et al.(1983)]{1983A&A...119L...3S} Shostak, G.~S., van Gorkom, J.~H., Ekers, R.~D., et al.\ 1983, \aap, 119, L3

\bibitem[Sugai et al.(2005)]{2005ApJ...629..131S} Sugai, H., Hattori, T., Kawai, A., et al.\ 2005, \apj, 629, 131%. doi:10.1086/431544

\bibitem[Vermeulen et al.(2003)]{2003A&A...401..113V} Vermeulen, R.~C., Ros, E., Kellermann, K.~I., et al.\ 2003, \aap, 401, 113%. doi:10.1051/0004-6361:20021752

\bibitem[Watson \& Wallin(1994)]{1994ApJ...432L..35W} Watson, W.~D. \& Wallin, B.~K.\ 1994, \apjl, 432, L35%. doi:10.1086/187505

\bibitem[Woo \& Urry(2002)]{2002ApJ...579..530W} Woo, J.-H. \& Urry, C.~M.\ 2002, \apj, 579, 530%. doi:10.1086/342878

\bibitem[Yamaki et al.(2012)]{2012PASJ...64..118Y} Yamaki, H., Kameno, S., Beppu, H., et al.\ 2012, \pasj, 64, 118%. doi:10.1093/pasj/64.5.118

\bibitem[Yates et al.(1997)]{1997MNRAS.285..303Y} Yates, J.~A., Field, D., \& Gray, M.~D.\ 1997, \mnras, 285, 303%. doi:10.1093/mnras/285.2.303

\end{thebibliography}
\end{document}